\documentclass[twocolumn]{aastex631}
\usepackage{amsmath}
\usepackage{graphics}
\usepackage{graphicx}
\usepackage{amssymb}
\usepackage{slashbox}

\begin{document}

\title{Enigmatic centi-SFU and mSFU nonthermal radio transients detected in the middle corona}

\correspondingauthor{Surajit Mondal}
\email{surajit.mondal@njit.edu}

\author[0000-0002-2325-5298]{Surajit Mondal}
\affiliation{Center for Solar-Terrestrial Research, New Jersey Institute of Technology, \\
323 M L King Jr Boulevard, Newark, NJ 07102-1982, USA}

\author[0000-0002-0660-3350]{Bin Chen}
\affiliation{Center for Solar-Terrestrial Research, New Jersey Institute of Technology, \\
323 M L King Jr Boulevard, Newark, NJ 07102-1982, USA}

\author[0000-0003-2872-2614]{Sijie Yu}
\affiliation{Center for Solar-Terrestrial Research, New Jersey Institute of Technology, \\
323 M L King Jr Boulevard, Newark, NJ 07102-1982, USA}

\author[0000-0002-1810-6706]{Xingyao Chen}
\affiliation{Center for Solar-Terrestrial Research, New Jersey Institute of Technology, \\
323 M L King Jr Boulevard, Newark, NJ 07102-1982, USA}

\author[0000-0001-6855-5799]{Peijin Zhang}
\affiliation{Center for Solar-Terrestrial Research, New Jersey Institute of Technology, \\
323 M L King Jr Boulevard, Newark, NJ 07102-1982, USA}
\affiliation{Cooperative Programs for the Advancement of Earth System Science, University Corporation for Atmospheric Research, Boulder, CO, USA}

\author{Dale Gary}
\affiliation{Center for Solar-Terrestrial Research, New Jersey Institute of Technology, \\
323 M L King Jr Boulevard, Newark, NJ 07102-1982, USA}

\author{Marin M. Anderson}
\affiliation{Owens Valley Radio Observatory, California Institute of Technology, Big Pine, CA 93513, USA}
\affiliation{Jet Propulsion Laboratory, California Institute of Technology, Pasadena, CA 91011, USA}
\author{Judd D. Bowman}
\affiliation{School of Earth and Space Exploration, Arizona State University, Tempe, AZ 85287, USA}
\author{Ruby Byrne}
\affiliation{Cahill Center for Astronomy and Astrophysics, California Institute of Technology, Pasadena, CA 91125, USA}
\affiliation{Owens Valley Radio Observatory, California Institute of Technology, Big Pine, CA 93513, USA}
\author{Morgan Catha}
\affiliation{Owens Valley Radio Observatory, California Institute of Technology, Big Pine, CA 93513, USA}
\author{Sherry Chhabra}
\affiliation{Center for Solar-Terrestrial Research, New Jersey Institute of Technology, \\
323 M L King Jr Boulevard, Newark, NJ 07102-1982, USA}
\affiliation{George Mason University, Fairfax, VA 22030, USA}
\author{Larry D'Addario}
\affiliation{Cahill Center for Astronomy and Astrophysics, California Institute of Technology, Pasadena, CA 91125, USA}
\affiliation{Owens Valley Radio Observatory, California Institute of Technology, Big Pine, CA 93513, USA}
\author{Ivey Davis}
\affiliation{Cahill Center for Astronomy and Astrophysics, California Institute of Technology, Pasadena, CA 91125, USA}
\affiliation{Owens Valley Radio Observatory, California Institute of Technology, Big Pine, CA 93513, USA}
\author{Jayce Dowell}
\affiliation{University of New Mexico, Albuquerque, NM 87131, USA}
\author{Gregg Hallinan}
\affiliation{Cahill Center for Astronomy and Astrophysics, California Institute of Technology, Pasadena, CA 91125, USA}
\affiliation{Owens Valley Radio Observatory, California Institute of Technology, Big Pine, CA 93513, USA}
\author{Charlie Harnach}
\affiliation{Owens Valley Radio Observatory, California Institute of Technology, Big Pine, CA 93513, USA}
\author{Greg Hellbourg}
\affiliation{Cahill Center for Astronomy and Astrophysics, California Institute of Technology, Pasadena, CA 91125, USA}
\affiliation{Owens Valley Radio Observatory, California Institute of Technology, Big Pine, CA 93513, USA}
\author{Jack Hickish}
\affiliation{Real-Time Radio Systems Ltd, Bournemouth, Dorset BH6 3LU, UK}
\author{Rick Hobbs}
\affiliation{Owens Valley Radio Observatory, California Institute of Technology, Big Pine, CA 93513, USA}
\author{David Hodge}
\affiliation{Cahill Center for Astronomy and Astrophysics, California Institute of Technology, Pasadena, CA 91125, USA}
\author{Mark Hodges}
\affiliation{Owens Valley Radio Observatory, California Institute of Technology, Big Pine, CA 93513, USA}
\author{Yuping Huang}
\affiliation{Cahill Center for Astronomy and Astrophysics, California Institute of Technology, Pasadena, CA 91125, USA}
\affiliation{Owens Valley Radio Observatory, California Institute of Technology, Big Pine, CA 93513, USA}
\author{Andrea Isella}
\affiliation{Department of Physics and Astronomy, Rice University, Houston, TX 77005, USA}
\author{Daniel C. Jacobs}
\affiliation{School of Earth and Space Exploration, Arizona State University, Tempe, AZ 85287, USA}
\author{Ghislain Kemby}
\affiliation{Owens Valley Radio Observatory, California Institute of Technology, Big Pine, CA 93513, USA}
\author{John T. Klinefelter}
\affiliation{Owens Valley Radio Observatory, California Institute of Technology, Big Pine, CA 93513, USA}
\author{Matthew Kolopanis}
\affiliation{School of Earth and Space Exploration, Arizona State University, Tempe, AZ 85287, USA}
\author{Nikita Kosogorov}
\affiliation{Cahill Center for Astronomy and Astrophysics, California Institute of Technology, Pasadena, CA 91125, USA}
\affiliation{Owens Valley Radio Observatory, California Institute of Technology, Big Pine, CA 93513, USA}
\author{James Lamb}
\affiliation{Owens Valley Radio Observatory, California Institute of Technology, Big Pine, CA 93513, USA}
\author{Casey Law}
\affiliation{Cahill Center for Astronomy and Astrophysics, California Institute of Technology, Pasadena, CA 91125, USA}
\affiliation{Owens Valley Radio Observatory, California Institute of Technology, Big Pine, CA 93513, USA}
\author{Nivedita Mahesh}
\affiliation{Cahill Center for Astronomy and Astrophysics, California Institute of Technology, Pasadena, CA 91125, USA}
\affiliation{Owens Valley Radio Observatory, California Institute of Technology, Big Pine, CA 93513, USA}
\author{Brian O'Donnell}
\affiliation{Center for Solar-Terrestrial Research, New Jersey Institute of Technology, \\
323 M L King Jr Boulevard, Newark, NJ 07102-1982, USA}
\author[0000-0001-6360-6972]{Kathryn Plant}
\affiliation{Owens Valley Radio Observatory, California Institute of Technology, Big Pine, CA 93513, USA}
\affiliation{Jet Propulsion Laboratory, California Institute of Technology, Pasadena, CA 91011, USA}
\author{Corey Posner}
\affiliation{Owens Valley Radio Observatory, California Institute of Technology, Big Pine, CA 93513, USA}
\author{Travis Powell}
\affiliation{Owens Valley Radio Observatory, California Institute of Technology, Big Pine, CA 93513, USA}
\author{Vinand Prayag}
\affiliation{Owens Valley Radio Observatory, California Institute of Technology, Big Pine, CA 93513, USA}
\author{Andres Rizo}
\affiliation{Owens Valley Radio Observatory, California Institute of Technology, Big Pine, CA 93513, USA}
\author{Andrew Romero-Wolf}
\affiliation{Jet Propulsion Laboratory, California Institute of Technology, Pasadena, CA 91011, USA}
\author{Jun Shi}
\affiliation{Cahill Center for Astronomy and Astrophysics, California Institute of Technology, Pasadena, CA 91125, USA}
\author{Greg Taylor}
\affiliation{University of New Mexico, Albuquerque, NM 87131, USA}
\author{Jordan Trim}
\affiliation{Owens Valley Radio Observatory, California Institute of Technology, Big Pine, CA 93513, USA}
\author{Mike Virgin}
\affiliation{Owens Valley Radio Observatory, California Institute of Technology, Big Pine, CA 93513, USA}
\author{Akshatha Vydula}
\affiliation{School of Earth and Space Exploration, Arizona State University, Tempe, AZ 85287, USA}
\author{Sandy Weinreb}
\affiliation{Cahill Center for Astronomy and Astrophysics, California Institute of Technology, Pasadena, CA 91125, USA}
\author{Scott White}
\affiliation{Owens Valley Radio Observatory, California Institute of Technology, Big Pine, CA 93513, USA}
\author{David Woody}
\affiliation{Owens Valley Radio Observatory, California Institute of Technology, Big Pine, CA 93513, USA}
\author{Thomas Zentmeyer}
\affiliation{Owens Valley Radio Observatory, California Institute of Technology, Big Pine, CA 93513, USA}

\begin{abstract}

Decades of solar coronal observations have provided substantial evidence for accelerated particles in the corona. In most cases, the location of particle acceleration can be roughly identified by combining high spatial and temporal resolution data from multiple instruments across a broad frequency range. In almost all cases, these nonthermal particles are associated with quiescent active regions, flares, and coronal mass ejections (CMEs). Only recently, some evidence of the existence of nonthermal electrons at locations outside these well-accepted regions has been found. Here, we report for the first time multiple cases of transient nonthermal emissions, in the heliocentric range of $\sim 3-7R_\odot$, which do not have any obvious counterparts in other wavebands, like white-light and extreme ultra-violet. These detections were made possible by the regular availability of high dynamic range low-frequency radio images from the Owens Valley Radio Observatory's Long Wavelength Array. While earlier detections of nonthermal emissions at these high heliocentric distances often had comparable extensions in the plane-of-sky, they were primarily been associated with radio CMEs, unlike the cases reported here. Thus, these results add on to the evidence that the middle corona is extremely dynamic and contains a population of nonthermal electrons, which is only becoming visible with high dynamic range low-frequency radio images. 

\end{abstract}

\section{Introduction}

The middle corona is defined as roughly the region of the corona spanning the heliocentric distances of 1.5--6$R_\odot$ \citep{west2023}. 
%It has a very tenuous plasma environment, where the density is too low for most X-ray and extreme ultraviolet (EUV) diagnostics. However, most studies of the low corona primarily use EUV and X-ray data. Due to this disparity in observation bands, which are sensitive to different characteristics of the plasma, along with their own limitations, it has been very quite hard in general to connect the inner and middle corona. This problem becomes particularly acute in the case of tracking outflowing structures like solar-wind blobs, waves etc. However, understanding and characterizing the middle corona is extremely important as this is the transition region between the highly disparate outer and inner corona and several key transitions in physical parameters occur in this region. Some of the crucial transitions which happen in the middle corona are the transition from plasma $\beta$ (defined as the ratio of magnetic pressure and thermal pressure) $<<1$ to $\beta >1$, transition from closed magnetic field configuration to fully open configuration, etc. 
Most studies of the middle corona are performed either using low-frequency radio observations or white light data from coronagraphs. Recently regular observations of the middle corona at EUV wavelengths are also being done \citep{darnel2022}. {The Extreme Ultraviolet Imager \citep{rochus2020}, onboard the SOlar Orbiter \citep{muller2020} can also provide EUV observations of the middle corona, when it is close to the Sun \citep{auchere2023}.} In the radio frequency range, frequencies approximately less than 300 MHz can directly probe the middle corona \citep{Chen2023decadal,west2023}. Although plasma parameters in the middle corona can be inferred in an indirect manner using observations at GHz frequencies \citep{kooi2014}, these require special observing conditions and are not suitable for regular observations and monitoring. { For example, Faraday rotation observations by \citet{kooi2014}, which measured the magnetic field in the middle corona, used data from an astrophysical linearly polarised source, when the line-of-sight to the source passes close to the Sun.} Solar observations in the meter wave frequency range have traditionally been done using spectrometers, which not only have poor sensitivity in general, but are also unable to distinguish between emissions happening at different locations on the Sun. The Nan\c{c}ay Radio Heliograph \citep[NRH,][]{bonmartin1983,avignon1989,kerdraon1997}, which can provide images and has been operating since the 2000s, has spot frequency coverages between 150 and 420 MHz. However, in the past decade, several low frequency radio interferometers have been commissioned, namely the Karl G. Jansky Very Large Array \citep[VLA,][]{perley2011}, Murchison Widefield Array \citep[MWA,][]{lonsdale2009, tingay2013, wayth2018}, and Low Frequency Radio Array \citep[LOFAR,][]{harlem13}, which not only has much higher sensitivity than older generation instruments, but is also able to provide images at very high time and frequency resolution. This new capability of time-frequency spectroscopy has opened up several new avenues, a review of which is provided in \citet{gary2023}. However, none of these instruments are dedicated to solar science, and hence regular monitoring observations of the Sun, often essential for studying the solar corona, are not possible with these instruments. The Owens Valley Radio Observatory's Long Wavelength Array (OVRO-LWA) is a low-frequency interferometer operating in the 13.4--86.9 MHz range (see \citealt{Anderson2018} for general descriptions of the instrument). It is an all-sky imager and consists of 352 dipoles spread over a 1.6 km $\times$ 2 km region after its most recent expansion. OVRO-LWA operates multiple observation modes simultaneously, which allows it to gather data suitable for a variety of science uses, including solar and space weather studies. It is considered a solar-dedicated instrument since it will be operating throughout the day and the Sun remains within its field of view throughout daylight hours\footnote{An overview paper describing the recently expanded OVRO-LWA and its solar capabilities is currently under preparation.}.
 %Also the Sun is dominant radio brightness in the daytime sky and its spatial resolution is sufficient to resolve key solar radio sources. 
%In addition, the OVRO-LWA has multiple specialized operating modes suitable for solar studies.
 
 To date, all of the radio emission reported in the literature beyond approximately 1.7-1.8 $R_\odot$ has either been coherent solar radio bursts, with brightness temperatures of several tens of MK \citep[e.g.][]{dulk1980,reid2017,chhabra2021,dabrowski2023} or ``radio CMEs'' \citep[e.g.;][]{bastian2001,maia2007,demoulin2012,mondal2020,kansabanik2023,kansabanik2024}\footnote{We use the term `radio CME' to imply extended radio emission having close resemblance with the structure of the white-light CME.}. 
% Contrary to past results, recently \citet{xingyao2025} reported radio emission from a CME with brightness temperature of several hundred MK, where the emission morphology roughly matched the CME structure. { Xingyao suggested that this goes against the radio CME having weak flux.}
 Here, we report the detection of multiple extremely weak radio transients observed between April 11--22, 2024 at large coronal heights ($\gtrsim 3 R_\odot$). The transients have a brightness temperature of only $10^4-10^5$K and are highly extended in nature, unlike the typical compact morphology expected from solar radio bursts. Interestingly, we do not detect any significant transient white light emission in the C2 detector of the Large Angle and Spectrometric Coronagraph Experiment (LASCO) onboard the Solar and Heliospheric Observatory (SOHO) at the location of these radio transients. This is unlike the radio CMEs, which show similar morphology to the emissions we report here. Thus, with its high sensitivity and capability to provide very high dynamic range solar images, OVRO-LWA has opened up an unexplored phase space in terms of the radio emissions observed in the middle corona. 

 This paper is organized as follows. Section \ref{sec:identification} describes the methodology we follow for the identification of these transients. Section \ref{sec:results} presents our results for these identified transients. In Section \ref{sec:discussion} we discuss the implications of these results, and present our conclusions in Section \ref{sec:conclusion}.

 \section{Identification of weak radio transients} \label{sec:identification}

The goal of this project was to identify and characterize faint emissions located high in the solar corona, beyond the solar limb. Although radio CMEs \citep[e.g.][]{bastian2001, mondal2020, kansabanik2023}, are a class of such emissions, our search algorithm does not enforce the presence of a CME during the observed radio emission. While the identification process is automated, each candidate is manually vetted for accuracy. We describe the algorithm below.

The algorithm follows two main steps:
\begin{enumerate}
    \item Identify the detected solar emission in a robust manner.
    \item If the extent of the radio emission is significantly larger than the expected size of the quiet sun at that frequency, the image is identified as a potential candidate and noted down for further manual vetting.
\end{enumerate}

To separate the ``true" sources from the noise, we have implemented a simple rms/standard deviation based thresholding, where we assume that most points which are above the $5\sigma$ threshold are real, where $\sigma$ is the rms of the image and is an estimate of the image noise. Generally, the rms is estimated from a sufficiently large source-free region. However, in this case, since the location and extent of the solar emission are unknown \textit{a priori}, we estimate the rms in a robust manner: we first find the rms considering all pixels in the image. Then we identify all pixels above the threshold and remove them before continuing to the next iteration of rms calculation and bright-pixel removal. This ensures that after some iterations, only pixels that are not related to the sources remain within the set and can be used to estimate the image noise in a robust manner. While this rms indeed gives an estimate of the noise, it is often found that close to the source, there are more artifacts, which can still lie above the threshold. While these can be removed by choosing a higher threshold, this would also mean that we will lose true source emission as well. Hence we do a clustering analysis using all the points above this threshold and choose only the largest group of points as consisting of real solar emission. { Unless the image is dynamic range limited (which can be determined from signal-to-noise considerations), the largest group is expected to be a patch containing the quiet sun disc and extended emissions out into the higher corona.} While it is true that the exclusion of smaller patches may lead to the removal of some genuine solar emissions in certain cases, we believe it is necessary to reduce false positives in our initial blind search for real sources.

\begin{figure*}[!ht]
    \centering
    \includegraphics[scale=0.75]{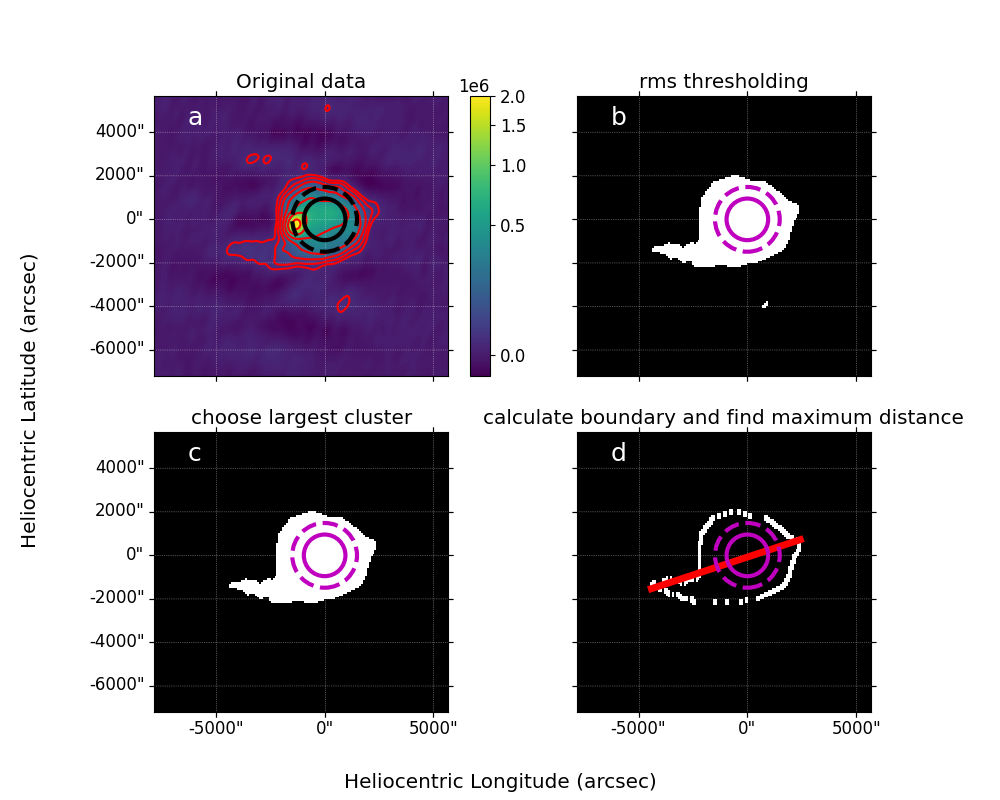}
    \caption{Shows the output of various steps involved in detecting a potential candidate. Data corresponds to 70 MHz at 16:57:37 UT on April 12, 2024. Panel a shows the original image. Contours of this image is also indicated to better show the extended structure towards the south east. Subsequent steps lead to panels b, c and d. {In Panel a, the lowest contour level is at 1.25\% of the peak and then increases in steps of 2. The dashed circle shows the size of the Sun at 70 MHz. The solid circle shows the optical disc of the Sun.}}
    \label{fig:cme_detection_steps}
\end{figure*}

After the solar emission is identified, we create a binary mask, where only the pixels where solar emission is detected are set to unity. We use an edge detection algorithm to determine the edge of the binary mask and then find the maximum distance between two points lying on the edge. This maximum distance provides us with the maximum extent of the solar radio emission. If the maximum distance is at least 1.5 times larger than the expected size of the quiet sun at that frequency, then we record that time as a potential candidate. The size of the quiet sun at each frequency is obtained from \citet{zhang2022}. The output of the various steps performed is shown for an example image at 16:57:37 at 70 MHz in Figure~\ref{fig:cme_detection_steps}. Panel (a) shows the original image in a linear color scale. Panel (b) shows the binary image created after applying the rms-based iterative thresholding. Panel (c) shows the result of choosing the largest cluster of source pixels. Please note that the small cluster of pixels around $(397\arcsec,-4000\arcsec)$, is not present in panel (c), but was present in panel (b). Panel (d) shows the result of the edge-detection algorithm. The points between which the largest distance was found have been joined using a red line. We have also provided some additional checks to remove spurious detections as well. We list below a few notable ones.

\begin{enumerate}
\item Radio images are always convolved with the angular resolution of the instrument. Hence, when the beam is very large, particularly when the solar elevation is small, the solar image appears artificially enlarged. To remove the effect of the beam, we ensure that the maximum distance detected is at least 3 times larger than the projection of the beam towards the direction of the maximum distance. If this condition is valid, we subtract the psf projection and use the subtracted quantity as the true maximum distance. {Apart from this, we restrict the search to times between 15:30--23:30 UT. Due to the low elevation of Sun outside this window, baselines along the East-West direction becomes particularly small, resulting in a very poor point-spread-function. }

%\item We also assume that if the distance of two points on the detected edge, lying perpendicular to the direction along which the maximum distance is detected, is comparable to the maximum distance, then it is most likely emission from quiet sun and hence will be ignored.

\item Once an extended emission, larger than the expected size of the quiet Sun at a given frequency, is detected, we also investigate if it is caused by ionospheric activity. At these low radio frequencies, ionosphere effects are quite important, and can even change the source morphology \citep[e.g.][]{boyde2022,dorrian2023}. We assume that the ionospheric effects are equally likely to increase the solar size, as it is likely to decrease the solar size, over a sufficiently long period of strong ionospheric activity. Hence to investigate if the extended emissions are likely to be caused by ionospheric activity, we investigated if within a span of 2 hours centered on the detected emission, we see evidence of a distorted quiet sun. If we find that the quiet sun disk is stable over this 2 hour period, we consider that the ionosphere is not active and the detected extended emission has a solar origin. We show the results of this investigation for the detected dates in Section \ref{sec:investigate_ionospheric_activity}.

\end{enumerate}

\section{Results} \label{sec:results} 

We applied the technique described in Section \ref{sec:identification} on all days between April 12--22, 2024. We used the images available from the OVRO-LWA database for this purpose. All images were generated by integrating over 5 MHz and 10s. {The cadence of images available in the database during the month of April 2024 is generally 1 minute, barring a few time periods where imaging was done at 10s cadence\footnote{These data are taken before completion of the instrument commissioning. During this time, some frames were dropped during imaging due to several factors like machine load, memory issues, IO issues, etc. These issues have been fixed now}.} The data processing was done by the OVRO-LWA solar realtime pipeline, which is described in detail in Mondal et al. (in preparation). { Here we only briefly describe the pipeline we have implemented for the data processing.

In the data analysis pipeline, we correct for both the instrumental effects as well as effects due to propagation of the radio waves through the ionosphere. Due to the simple architecture of OVRO-LWA, the instrument is very stable, and hence instrumental effects are typically calibrated a few times a month. For this purpose, we generally use nighttime data, when bright astronomical sources like Cygnus A, Cassiopeia A, Taurus A, or Virgo A are in the sky. Any changes in the antenna gains from the values determined from the nighttime calibration data are corrected using self-calibration techniques. We then apply the antenna gains derived from night-time calibration data, as well as those obtained through self-calibration, to determine the final calibrated data. Next, we subtract bright astronomical sources present in the sky, and change the phase center to the solar location. We use these data to produce images with an FOV of $96 R_\odot$ at both 384 kHz and 5 MHz frequency resolution. A central image of size $\approx24R_\odot$ is cropped and saved in the long-term data archive.

We made multiple detections of extended emission beyond the solar limb between April 12-22, 2024. }Most of the detections were associated with CMEs. However, in some instances, we also found that there was no significant or very weak white light transient emission at the location where we detected a radio transient. Here, we focus on such transient radio emissions. In the following subsections, we present these detections and their properties.

\subsection{Events on 2024 April 12}

\begin{figure*}
    \centering
    \includegraphics[trim={0.5cm 0 0 0},clip,scale=0.45]{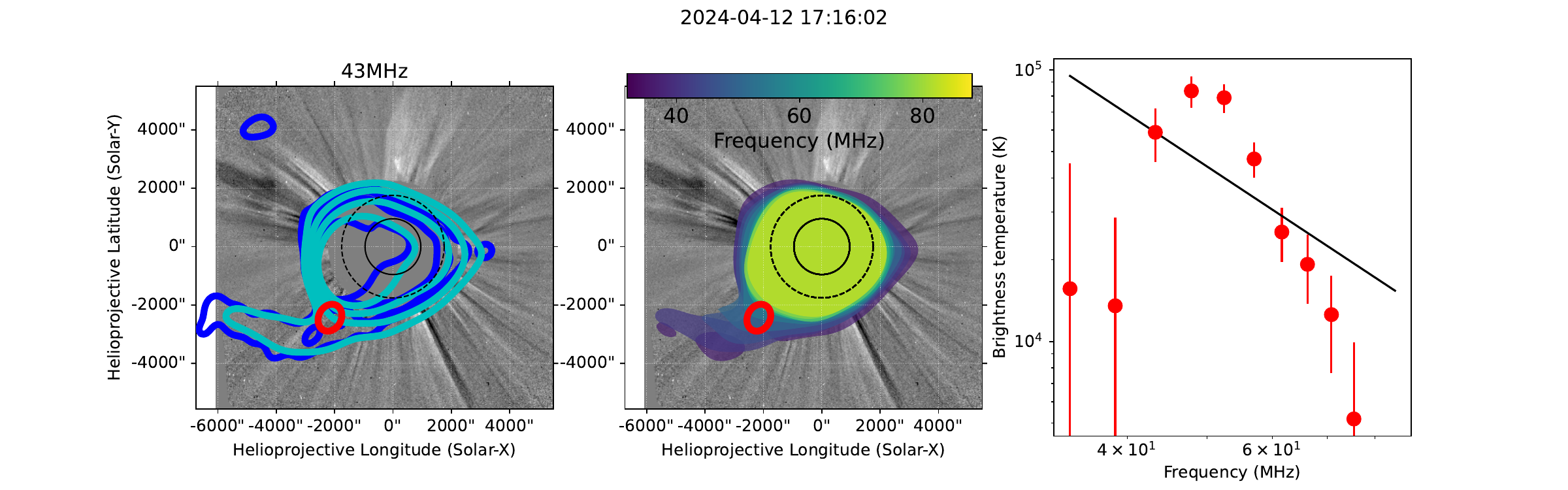}
    \caption{Left panel: Contours at 43 MHz and 17:16 on April 12, 2024, are overlaid on the nearest LASCO C2 difference image. Difference has been taken with respect to a 2-hour prior image. The lowest contour is at 0.04 MK and then increases in multiples of 2. Middle panel: Multiple frequency radio contours are overlaid on the same LASCO C2 difference image. The outer contour level at each frequency is at 0.04 MK. Black dashed circle in the left and middle panels shows the radio sun at 43 MHz. The solid circles show the optical disc of the Sun. Right panel: Shows the spectrum extracted from the region marked in the left panel. {The spectrum has been extracted from the original image before any smoothing was done.} Circles have been used to denote detections. The black line shows a spectrum following $\nu^{-2}$ dependence, where $\nu$ is the observation frequency. {All radio contours, except the ones shown in the left panel using blue color, have been obtained by smoothing the images using a Gaussian function with standard deviation of $7.5\arcmin$.} }
    \label{fig:20240412_event}
\end{figure*}

In the left panel of Figure \ref{fig:20240412_event}, we overlay the contours of the OVRO-LWA 43 MHz image at 17:16 UT over the closest LASCO C2 difference white light image (taken with respect to the image 2 hours before). The outermost contour level is at 0.04 MK and then increases in multiples of 2 (i.e., 0.04 MK, 0.08 MK, 0.16 MK, and 0.32 MK). { The original image is shown using blue contours. We see an extended radio emission from the south-eastern limb, and the structure extends from $\sim 3.3-6.5R_\odot$. It can also be seen that there are some noise contours as well, at the same brightness temperature level as the faint extended emission.} To {increase the fidelity of} the extended emission better, we have smoothed the image at each frequency by a Gaussian function with a standard deviation of $7.5^{'}$.{ In comparison, the native resolution of the image at 43 MHz can be described by a Gaussian with standard deviation of $6.8\arcmin \times 5.4 \arcmin$. { After smoothing, the effective resolution of the same image increased to that of a Gaussian with standard deviation of $10.1\arcmin \times 9.3\arcmin$. The contours of the smoothed image is shown using cyan. No noise peak at a similar level as the extended emission is observed in the smoothed image.}
%We have also plotted a negative contour at $-$0.04 MK using a dashed line. Absence of negative contours is often used to show the image fidelity, and the fact that negative contours are not present demonstrates that the image is reliable at a brightness temperature level greater than 0.04 MK.}
In the middle panel, we show the radio contours of the smoothed images at multiple frequencies. 
The outermost contour is again at 0.04 MK. {The black dashed circles in the left and right panels correspond to the size of the radio Sun at 43 MHz \citep{zhang2022}.} 
%The high fidelity of the contours is self-evident.  
In the right panel, we show the spectrum extracted from the region marked in the left panel. The shape and size of this region is equal to the instrumental resolution at the lowest frequency shown in the spectrum in the right panel. {The spectrum has been extracted using the unsmoothed original images.} In the right panel, we have used circles to denote the detections. We define detections as the frequencies where the peak brightness temperature value within each region is greater than 5$\sigma$. For frequencies with detections, we have plotted a circle. { The 1$\sigma$ error bars shown in the spectra are set to be the rms value of a source-free region in the image.} The black lines show the expected spectra if the spectra were dominated by optically thin free-free emission in the absence of magnetic field.

{In the left panel of Figure \ref{fig:total_flux_20240412}, we have shown the difference between the 43 MHz radio images at 17:16:00 and 17:03:28 UT. The difference image has been smoothed using a Gaussian function with a standard deviation of $7.5\arcmin$ to bring out the faint extended emission. The contour has been drawn at 0.03 MK. The black dashed line indicates the extent of the transient emission at 43 MHz. In the right panel of Figure \ref{fig:total_flux_20240412}, we show the total flux of the transient. The area where the transient emission is detected was chosen manually. Care was taken to exclude any emission from the quiet sun, which can also show some temporal variability. {Pixels, which are both within the black dashed line, as well as have a plane-of-sky heliocentric distance smaller than the size of the quiet sun at 43 MHz, have been excluded.} The triangles indicate the $5\sigma$ upper limits, and are shown where the transient is not detected with a signal-to-noise ratio (SNR) of 3. We have used a lower SNR threshold as higher frequencies generally occupy a region much smaller than the area chosen for flux computation, and hence can have lower SNR in spite of having genuine transient emission.}
\begin{figure*}
    \centering
    \includegraphics[scale=0.5]{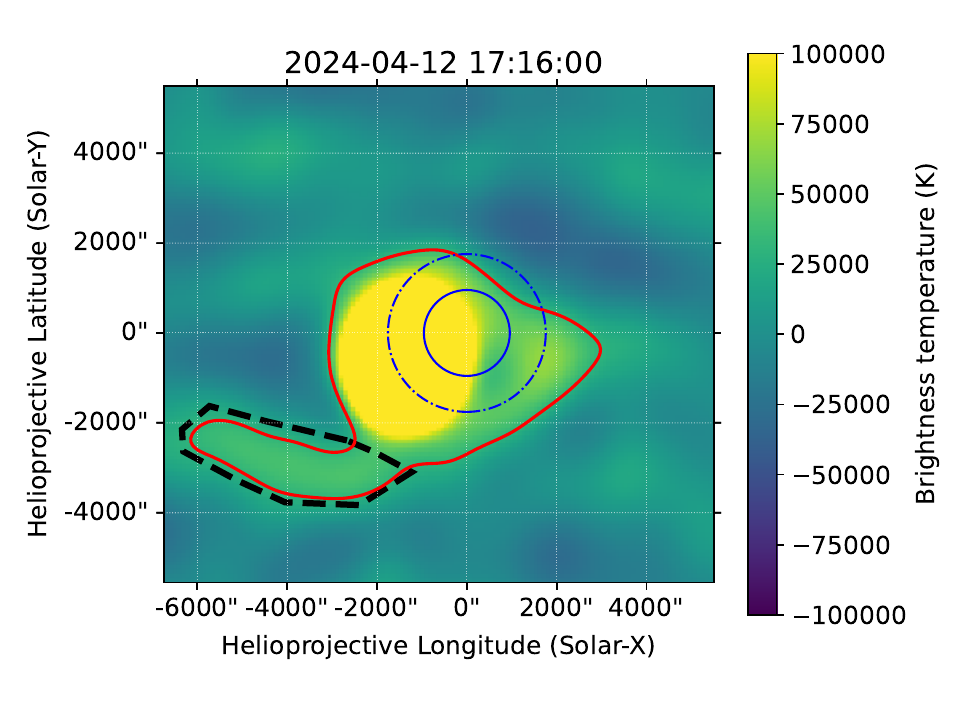}
    \includegraphics[scale=0.5]{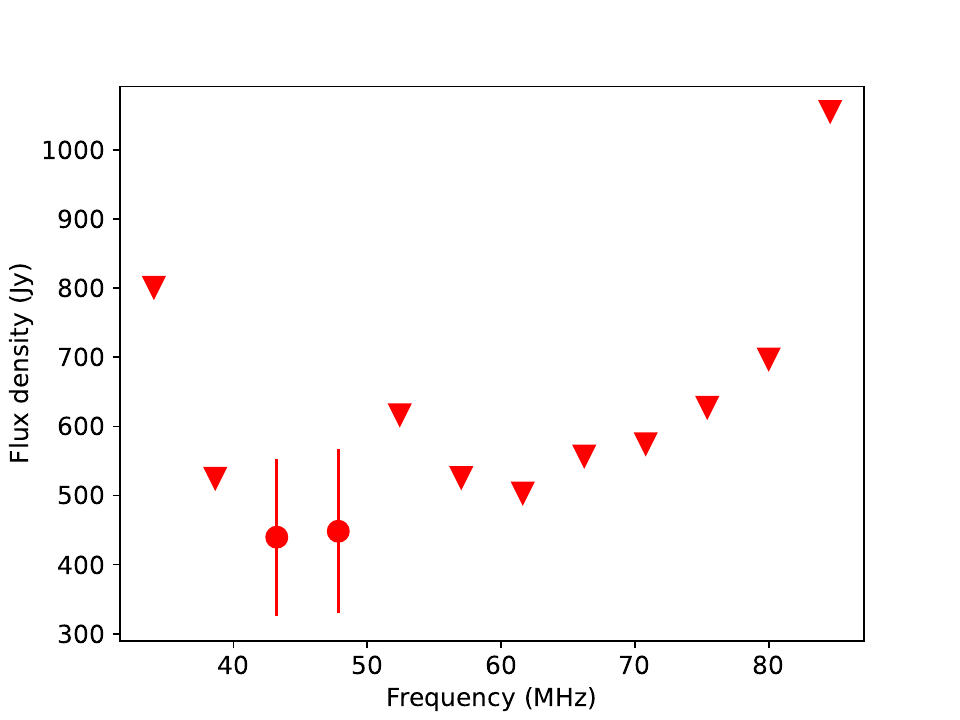}
    \caption{Left panel: Difference 43 MHz radio image between 17:16:00 and 17:03:28 UT. The red contour shows the 0.03 MK. The black dashed line indicates the extent of the transient emission at 43 MHz. The solid and dashed circles show the solar limb in the optical and 43 MHz images, respectively. Right panel: Shows the total flux of the transient. The triangles indicate the $5\sigma$ upper limits.}
    \label{fig:total_flux_20240412}
\end{figure*}

\begin{figure*}
    \includegraphics[scale=0.5]{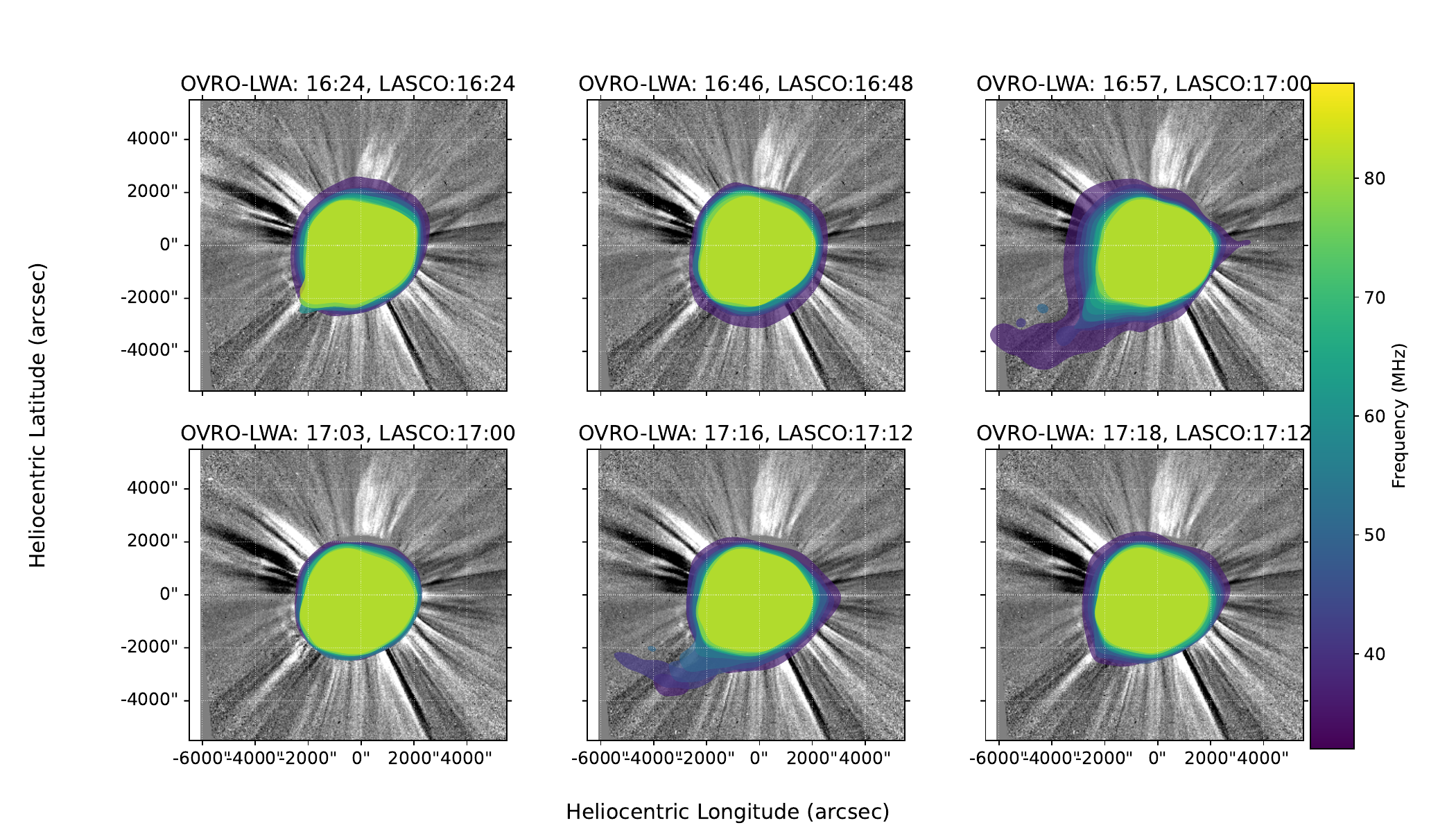}
    \caption{Multi-frequency contours at different times overlaid on the nearest available LASCO C2 difference image. Difference has been taken with respect to LASCO C2 image at 14:18 UT. The lowest contour at each frequency and time is at 0.05 MK. }
    \label{fig:20240412_time_variability}
\end{figure*}

Figure~\ref{fig:20240412_time_variability} shows multi-frequency contours at different times overlaid on the nearest available LASCO C2 difference white light image (taken with respect to the LASCO C2 image at 14:18 UT). All radio images have been smoothed using a Gaussian of standard deviation equal to $4.5\arcmin$. The lowest contour at each frequency and time is at 0.05 MK. {A movie showing the temporal variability of the radio emission is available online in Figure \ref{fig:20240412_movie}.} It is evident from these figures, that the extended radio emission seen towards the eastern limb is transient in nature. The maximum extent of the extended emission is observed at 16:57 UT, when the emission extends to $\sim 7.5 R_\odot$. {Accurate estimation of the emission lifetime in each instance the emission was detected is problematic as the images are not available at a regular 1 minute cadence on this day\footnote{We define lifetime as the time duration for which an transient emission is continuously (within the limitations of the cadence of available images) detected.}. However we can put an upper limit on the lifetime, assuming the lifetime is comparable for all instances, using the fact that the emission was detected at 17:16 UT, but is not detected at 17:18 UT. Hence the lifetime of the emission is about 2 minutes. }
%While there is an indication of the extended emission towards the eastern limb at 16:24 UT, the extent of the extension is not as large as that observed in 16:57 and 17:16. 

\subsection{Events on 2024 April 14}

\begin{figure*}
    \centering
    \includegraphics[trim={0.5cm 0 0 0},clip,scale=0.45]{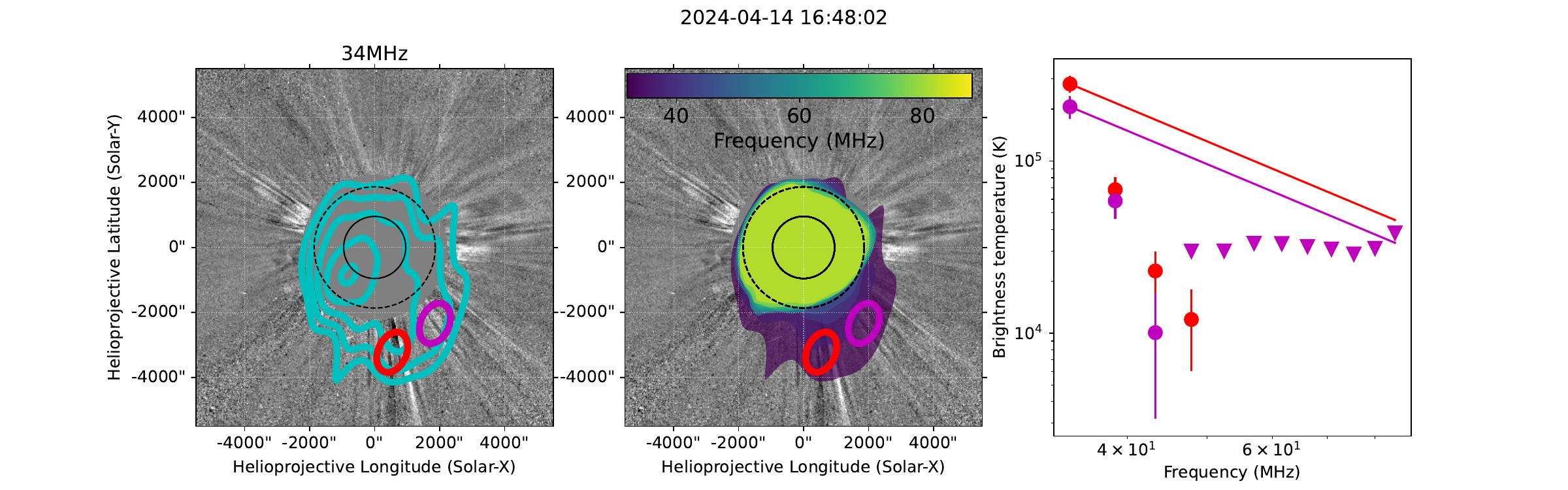}
    \includegraphics[trim={0.2cm 0 4cm 0},clip,scale=0.45]{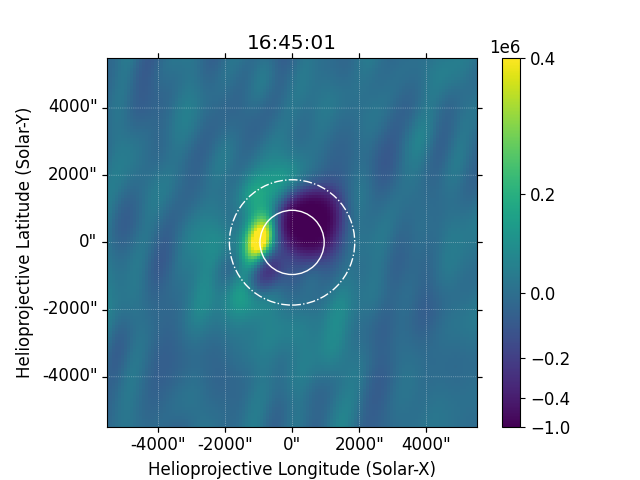}
    \includegraphics[trim={0.77cm 0 4cm 0},clip,scale=0.45]{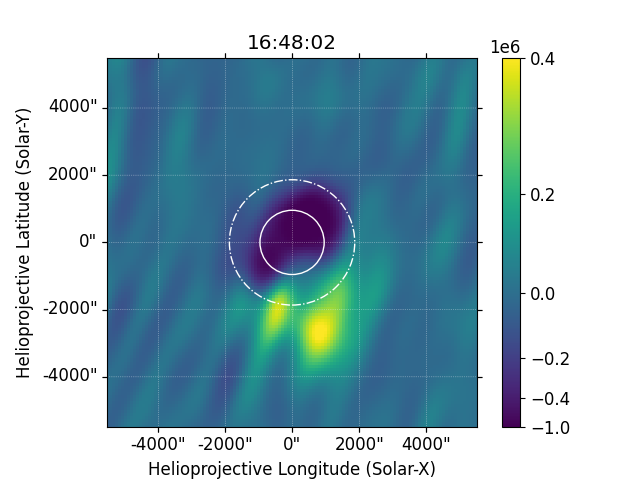}
    \includegraphics[trim={0.79cm 0 1cm 0},clip,scale=0.45]{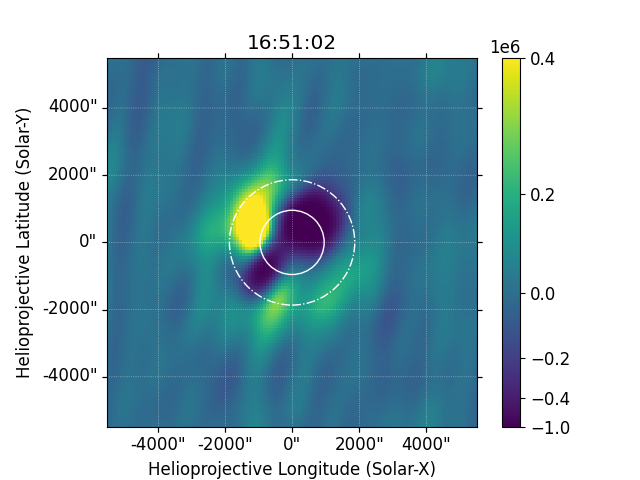}
    \caption{Top left panel: Radio contours at 34 MHz and 16:48 on April 14, 2024 are overlaid on the nearest LASCO C2 difference image. Difference has been taken with respect to a LASCO C2 image 1 hour prior to this time. The lowest contour is at 0.08 MK and then increases in multiples of 2. 
    {Negative contour at -0.08 MK (corresponding to lowest positive contour) is drawn using a cyan dashed line.} {The fact that no dashed line is visible, indicates the absence of any negative peak whose absolute value exceeds or is equal to 0.08 MK. } 
    Top middle panel: Multiple frequency radio contours are overlaid on the same LASCO C2 difference image. The outer contour level at each frequency is at 0.08 MK. {The black dashed line in left and middle panel shows the radio sun at 34 MHz.} The solid circles show the optical disc of the Sun. Top right panel: Brightness temperature spectrum of 2 regions shown. The marker color used to show the spectrum of each region is same as that used to indicate the location of the corresponding region in the top left panel. Properties of the markers and colors are the same as those in Figure \ref{fig:20240412_event}. Triangles have been used to denote the non-detections and represent the 5$\sigma$ upper limit. The red and magenta lines show the spectra following $\nu^{-2}$ dependence, where $\nu$ is the observation frequency.  Bottom panels: Difference 34 MHz images for 16:45, 16:48 and 16:51 UT are shown in the bottom left, bottom middle and bottom right panels respectively. Difference has been taken with respect to the image at 16:32. The colorbar is highly saturated and non-linear and is the same for the 3 images.  }
    \label{fig:20240414_event}
\end{figure*}

In the top left panel of Figure \ref{fig:20240414_event}, we overlay the contour of the 34 MHz image obtained by OVRO-LWA on 2024 April 14 at 16:48:02 UT over a LASCO C2 difference image, where the difference was taken with respect to the image 1 hour before this time of this image. The lowest contour level is at 0.08 MK and then increases in multiples of 2. We observe the presence of an extended emission towards the south between $\sim 2.2-4.2 R_\odot$. The top middle panel shows multi-frequency contours at 0.08 MK overlaid on the same LASCO C2 difference image. {The black dashed circles in the top left and right panels corresponds to the size of the radio Sun at 34 MHz \citep{zhang2022}.} On the top right panel, we show the spectra extracted from the two regions shown in the top left panel. The color of markers used in the spectrum is the same as that corresponding region shown in the left panel. Other properties of the image are the same as those of Figure \ref{fig:20240412_event}. For non-detections, we have put a triangle at the $5\sigma$ level, which indicates the upper limit to the flux at that frequency. The spectrum of both regions shows a steep decrease in flux density with an increase in frequency. Bottom panel shows the 34 MHz difference images at 16:45, 16:48, and 16:51 UT, where the difference has been taken with respect to the image at 16:32.  The same non-linear highly saturated color scale has been used to show the three images. It is evident that the extended emission seen towards the south at 16:48 UT is not present in the images at 16:45 UT and 16:51 UT, in spite of very similar noise characteristics to the 16:48 UT image. In fact, we find that this extended emission is only detected at 16:47 and 16:48 UT. This demonstrates that this emission is highly transient and has a lifetime of the order of a few minutes. The LASCO difference image, on the other hand, does not show any signature that is similar in extent to that seen in the radio data.

{In the left panel of Figure \ref{fig:total_flux_20240414}, we have shown the difference between the 34 MHz radio images at 16:48:02 and 16:32:09 UT. The contour has been drawn at 0.16 MK. The black dashed line indicates the extent of the transient emission at 34 MHz. In the right panel of Figure \ref{fig:total_flux_20240414} we show the total flux of the transient. { Similar to Figure \ref{fig:total_flux_20240412}, points, which are both within the black dashed line, as well as have a plane-of-sky heliocentric distance smaller than the size of the quiet sun at 34 MHz, have been excluded.} The triangles indicate the $5\sigma$ upper limits, and is shown where the transient is not detected with a SNR of 3.}
\begin{figure*}
    \centering
    \includegraphics[scale=0.5]{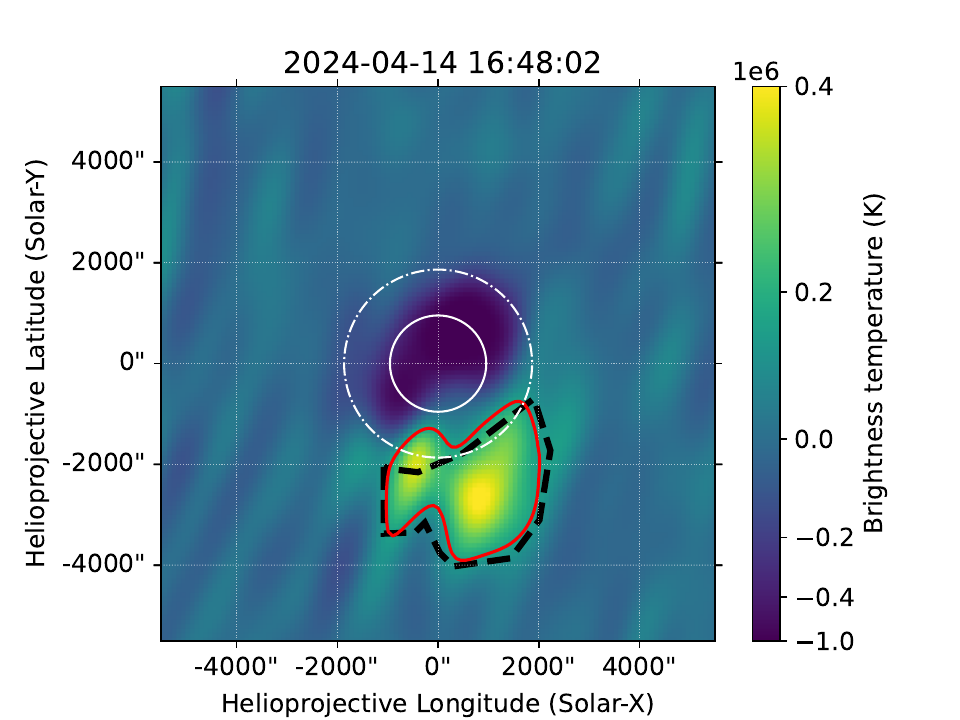}
    \includegraphics[scale=0.5]{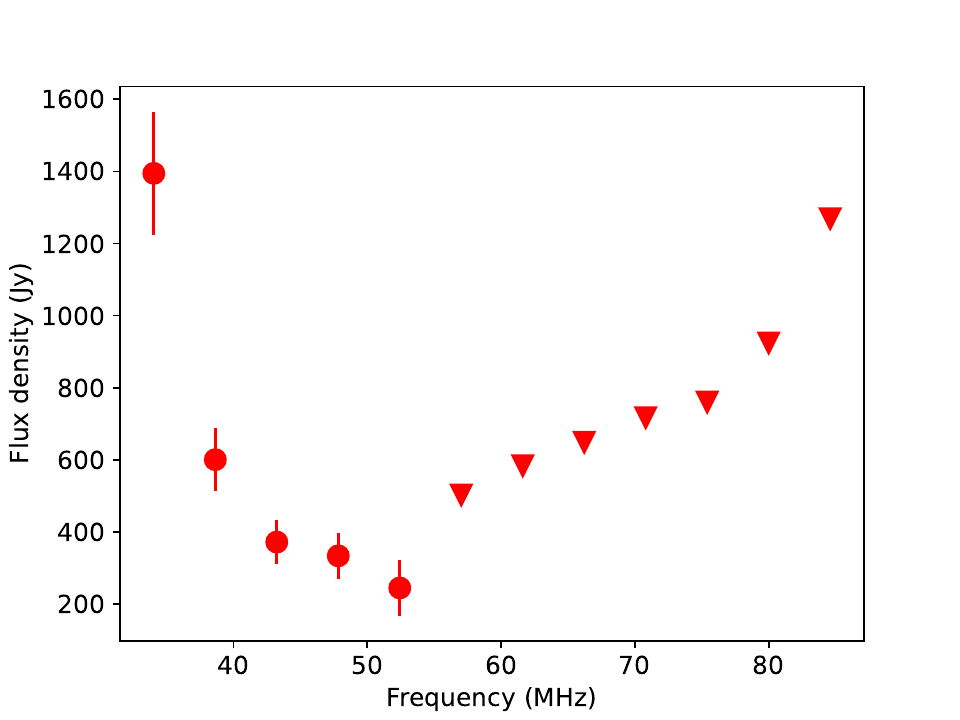}
    \caption{Left panel: Shows the difference of 34 MHz radio image at 16:48:02 and 16:32:09 UT. The contour has been drawn at 0.16 MK. The black dashed line indicates the extent of the transient emission at 34 MHz. The solid and dashed circles show the limb of the Sun in the optical and 34 MHz images respectively. Right panel: Shows the total flux of the transient. The triangles indicate the $5\sigma$ upper limits.}
    \label{fig:total_flux_20240414}
\end{figure*}

\subsection{Events on 2024 April 22}

\begin{figure*}
    \centering
    \includegraphics[trim={0.5cm 0 0 0},clip,scale=0.45]{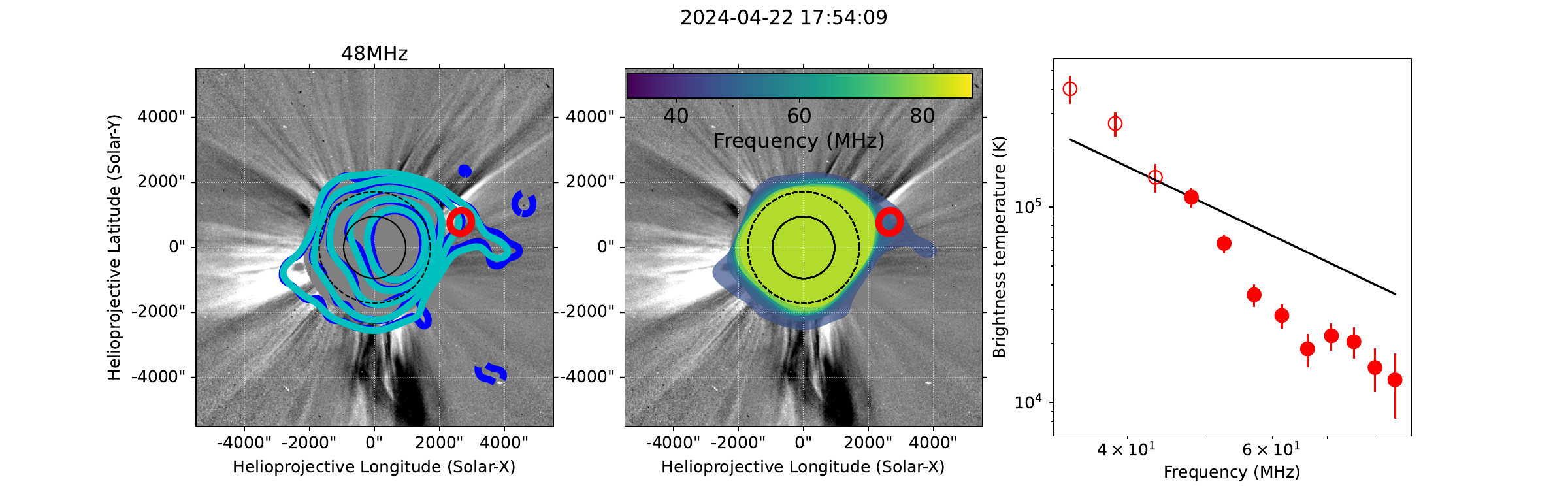}
    \caption{Left panel: 0.06 MK contour at 48 MHz and 17:54 on April 22, 2024, is overlaid on the nearest LASCO C2 difference image. Difference has been taken with respect to a 4 hour prior image. {Negative contour at -0.06 MK (corresponding to lowest positive contour) is also shown using dashed line.} { Blue and cyan contours have been used to show the original and smoothed version of the image. Smoothing was done by a Gaussian with a standard deviation of $4.5^{'}$ to bring out the faint emission better.The fact that no cyan dashed line is visible, indicates the absence of any negative peak in the smoothed image whose absolute value exceeds or is equal to 0.06 MK. This demonstrates the high fidelity of the detected extended emission.}  Middle panel: Multiple frequency radio contours {of the smoothed images} are overlaid on the same LASCO C2 difference image. The outer contour level at each frequency is at 0.06 MK. Black dashed circle in the left and middle panels show the radio sun at 48 MHz. The solid circle shows the optical disc of the Sun. Other properties of the images are same as that in Figure \ref{fig:20240412_event}. In the right panel, we have used hollow symbols to show the flux densities measured at 34, 39 and 43 MHz, as they are affected by a bright source located close-by.}
    \label{fig:20240422_event}
\end{figure*}

In the left panel of Figure \ref{fig:20240422_event}, we have overlaid the OVRO-LWA radio contours at 48 MHz from 2024 April 22 at 17:54 UT on a LASCO C2 difference image {using blue}. The difference has been taken with respect to a 4-hour prior image.  The lowest contour is at 0.06 MK, and then increases in steps of 2. Negative contour at -0.06 MK (shown using dashed line) is also shown. Negative contours are clearly visible. To increase the fidelity of the faint extended emission better, the original image available in the database has been smoothed by a Gaussian with a standard deviation of $4.5^{'}$. { In comparison, the native resolution of the image at 48 MHz can be described by a Gaussian function with a standard deviation of $6.3\arcmin \times 5.7 \arcmin$. } { After smoothing, the effective resolution of the same image increased to that of a Gaussian with standard deviation of $7.7\arcmin \times 7.3\arcmin$. Contours of the smoothed image are shown using cyan. Contour levels are same as that of the original image. It is evident that no noise peak or negative contours are observed at the level of extended emission after smoothing. }On the middle panel, we have overlaid radio contours of {the smoothed} images at multiple frequencies of the same LASCO C2 difference image. The lowest contour is again at 0.06 MK, and similar to the 43 MHz image shown in the left panel, images at all frequencies have also been smoothed by the same smoothing kernel.  Additionally, we have not shown contours at the 34, 39, and 43 MHz due to dynamic range limitations. {The black dashed circles in the top left and right panels correspond to the size of the radio Sun at 48 MHz \citep{zhang2022}.} We observe an extended emission from the western limb in both the left and middle panels. This emission extended from about $2.2-4 R_\odot$. While the extension is most clearly visible in 48 MHz, 52 MHz also shows an extension, although, it is much less extended compared to the 48 MHz image. While there is a white light structure close to this radio extension, based on the direction of the emission, we think that the radio emission is not related to the white-light structure.  In the right panel, we have shown the spectra extracted from the region shown in the left panel. The brightness temperature shown for 34, 39, and 43 MHz is affected by a bright source located close by and hence has been shown using hollow symbols, unlike the filled symbols used for other frequencies. All other properties of the image are the same as those in Figure \ref{fig:20240412_event}. 

{In the left panel of Figure \ref{fig:total_flux_20240422}, we have shown the difference between the 48 MHz radio images at 17:54:09 and 17:51:08 UT. The difference image has been smoothed using a Gaussian function with a standard deviation of $4.5\arcmin$ to bring out the faint extended emission.  The contour has been drawn at 0.05 MK. The black dashed line indicates the extent of the transient emission at 48 MHz. In the right panel of Figure \ref{fig:total_flux_20240414}, we show the total flux of the transient. { Similar to Figure \ref{fig:total_flux_20240412}, pixels, which are both within the black dashed line, as well as have a plane-of-sky heliocentric distance smaller than the size of the quiet sun at 48 MHz, have been excluded.} The triangles indicate the $5\sigma$ upper limits, and are shown where the transient is not detected with a SNR of 3.}
\begin{figure*}
    \centering
    \includegraphics[scale=0.5]{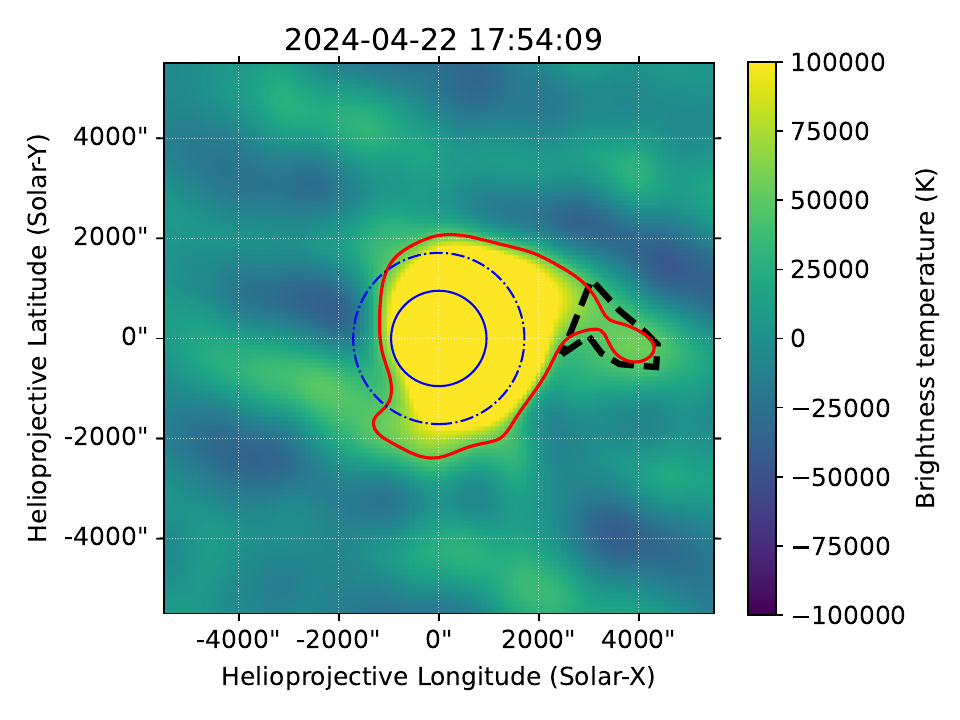}
    \includegraphics[scale=0.5]{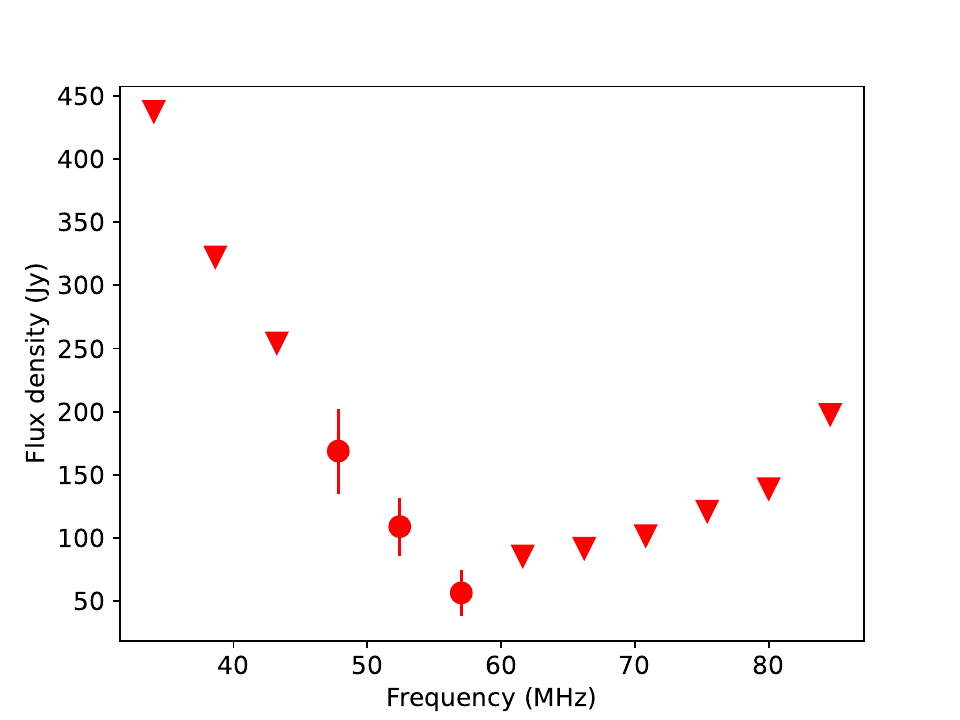}
    \caption{Left panel: Difference 48 MHz radio image between 17:54:09 and 17:51:08 UT. The contour has been drawn at 0.05 MK. The black dashed line indicates the extent of the transient emission at 48 MHz. The solid and dashed circles show the solar limb in the optical and 48 MHz images, respectively. Right panel: Shows the total flux of the transient. The triangles indicate the $5\sigma$ upper limits.}
    \label{fig:total_flux_20240422}
\end{figure*}

Figure \ref{fig:20240422_time_variability} shows multi-frequency contours at different times overlaid on the nearest available LASCO C2 difference white light image. The difference has been taken with respect to LASCO C2 image at 14:18 UT. All radio images have been smoothed using a Gaussian of standard deviation equal to $4.5^{'}$. The lowest contour at each frequency and time is at 0.05 MK. We have not shown images that have very high noise. First, we note that a CME is visible in the LASCO C2 field of view at 16:57. Its signature is seen both towards the eastern and southern limb. {A movie showing the temporal variability of the radio emission is available online in Figure \ref{fig:20240422_movie}.} While we have detected the radio emission from the CME at multiple times, here we have only shown the detection at 16:57 and 17:21 UT. We will not discuss radio emission from CMEs here. At 15:50, we also see an extension from approximately the same location as that seen at 17:54 UT. In the intervening 2 hours, we find some indication of such extended emission, but the signal-to-noise-ratio of the detected sources were not high. {We also see that the white-light emission seen towards the north-western limb, close to the location of the extended radio source at 17:54 UT, is observed consistently between 15:36 and 18:00 UT. However the radio source is only detected at a few time instances in this interval. This bolsters our hypothesis that the extended radio emission detected here is not directly related to the white-light extension seen towards the north-western limb.} { Based on the Figure \ref{fig:20240422_movie} we again estimate the lifetime of the radio emission to be approximately a few minutes (1--2 minutes).}

\begin{figure*}
    \includegraphics[scale=0.7]{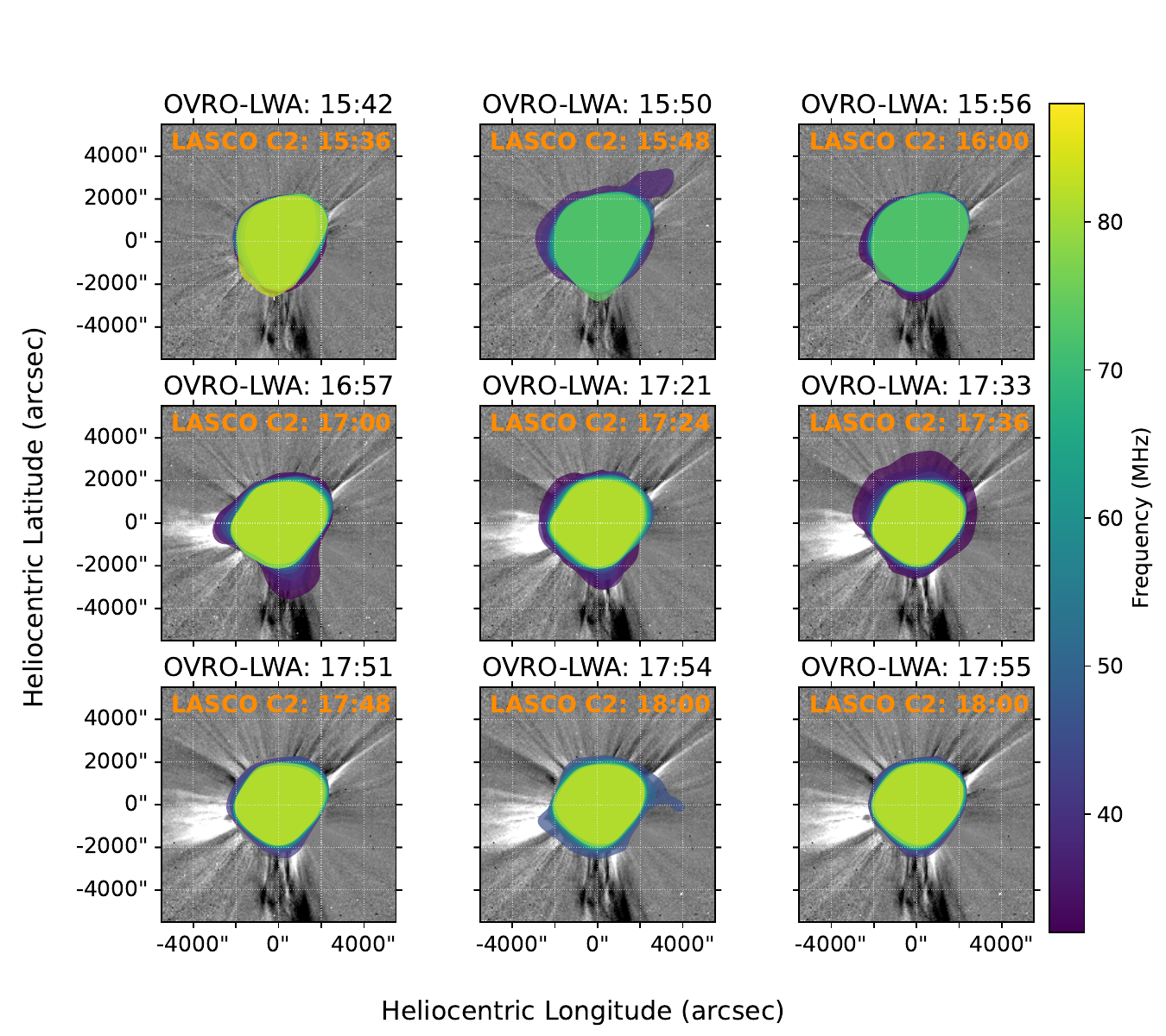}
    \caption{Multi-frequency contours at different times overlaid on the nearest available LASCO C2 difference image. Difference has been taken with respect to LASCO C2 image at 14:18 UT. The lowest contour at each frequency and time is at 0.05 MK. }
    \label{fig:20240422_time_variability}
\end{figure*}

% \begin{figure*}
%     \centering
%     \includegraphics[width=0.32\linewidth]{ovro-lwa.lev1.5_mfs_10s.2024-04-22T175209Z.image_I.png}
%     \includegraphics[width=0.32\linewidth]{ovro-lwa.lev1.5_mfs_10s.2024-04-22T175309Z.image_I.png}
%     \includegraphics[width=0.32\linewidth]{ovro-lwa.lev1.5_mfs_10s.2024-04-22T175509Z.image_I.png}
%     \caption{Caption}
%     \label{fig:20240422_multi_freq_overlay_time_series}
% \end{figure*}

% \begin{figure*}
%     \includegraphics[width=\linewidth]{ovro_lwa_overlay_timeseries_april22.png}
%     \caption{XXX}
%     \label{fig:april22_15UT_timeseries}
% \end{figure*}

\section{Discussion} \label{sec:discussion}

In Figures \ref{fig:20240412_event}, \ref{fig:20240414_event} and \ref{fig:20240422_event}, we show instances where we have detected extended transient radio emission. In Section \ref{sec:investigate_ionospheric_activity}, we discuss our tests evaluating the ionospheric activity on the days. In Section \ref{sec:emission_mechanism}, we try to determine the emission mechanisms which might be responsible for these emission structures.

\subsection{Impact of ionospheric activity on solar morphology} \label{sec:investigate_ionospheric_activity}

In Section \ref{sec:identification}, we discussed the methodology we follow to identify if the detected extended emissions can arise from strong ionospheric activity. We manually checked the images for each time instance for a 2 hour period centered on the event times as shown in Figures \ref{fig:20240412_event}, \ref{fig:20240414_event} and \ref{fig:20240422_event}, and verified that the quiet sun disc is undistorted in all images. For ease of visualization, we have created some 1D cuts which are shown in Figure \ref{fig:ionospheric_activity}. The figure is divided into 4 panels, corresponding to April 12, April 14, April 22 and April 11, 2024, from top left in a clockwise manner. Each panel is further subdivided into 2 subpanels. April 11, 2024, is shown for comparison purpose only, to demonstrate the effect of strong ionospheric activity. For April 11, we have shown a smaller time interval for better visualization of the quiet sun artefacts observed in presence of strong ionospheric activity. For uniformity, we have shown results from the 43 MHz image at all days. We have verified that for April 14, 34 MHz image also shows similar results. A black horizontal dashed line shows a brightness temperature of 0.1 MK, and is an arbitrary threshold chosen to demarcate the quiet solar disc. This threshold was also chosen by \citet{zhang2022} to determine the size of the quiet Sun at these frequencies. The left and right subpanels show the 1D brightness temperature distribution along a line passing through the center of the Sun along the heliocentric longitude and latitude respectively. We have only shown the 1D cuts for figures for which the peak brightness temperature in the image is smaller than 2 MK, to remove dynamic range limited images. It is evident that for April 12, 14 and 22, all the 1D cuts have a diameter $\sim 64^{'}$, and it is approximately the same for all the 1D cuts along that direction. However, for April 11, the active ionosphere day, we find multiple instances where the quiet sun disc is much smaller, sometimes even becoming almost half the expected size. Based on this, we have ruled out strong ionospheric activity as the primary cause of the observed extended emissions.

\begin{figure*}
    \centering
    \includegraphics[scale=0.35]{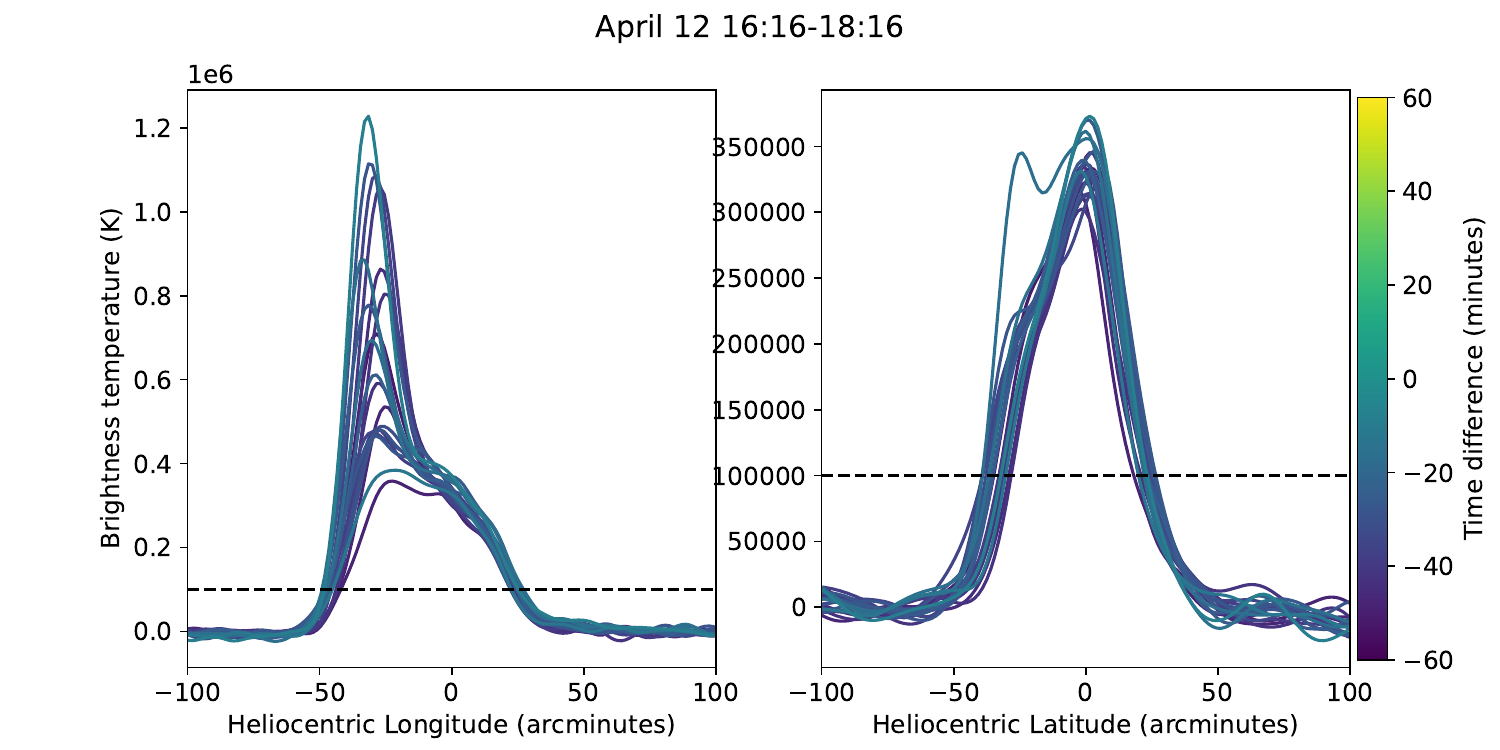}
    \includegraphics[scale=0.35]{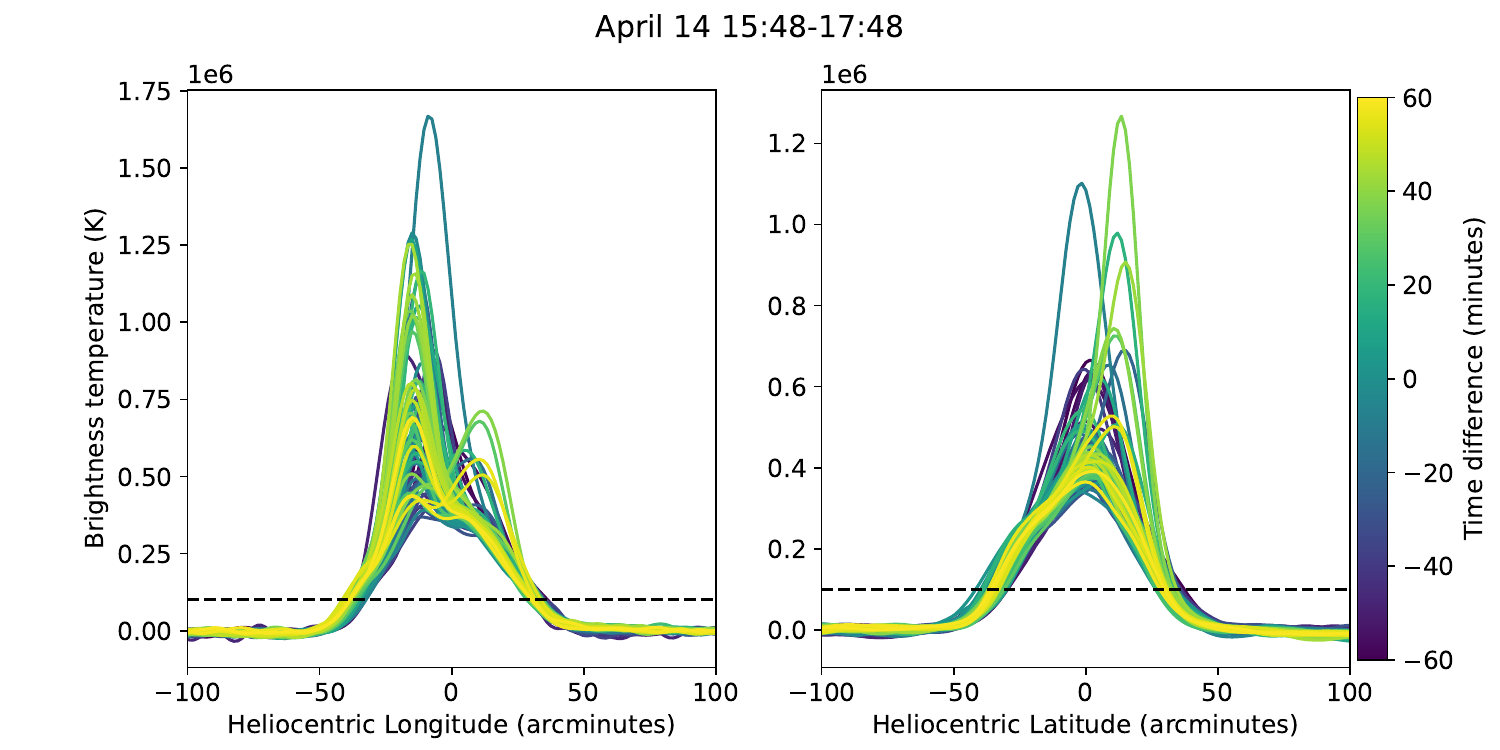}
    \includegraphics[scale=0.35]{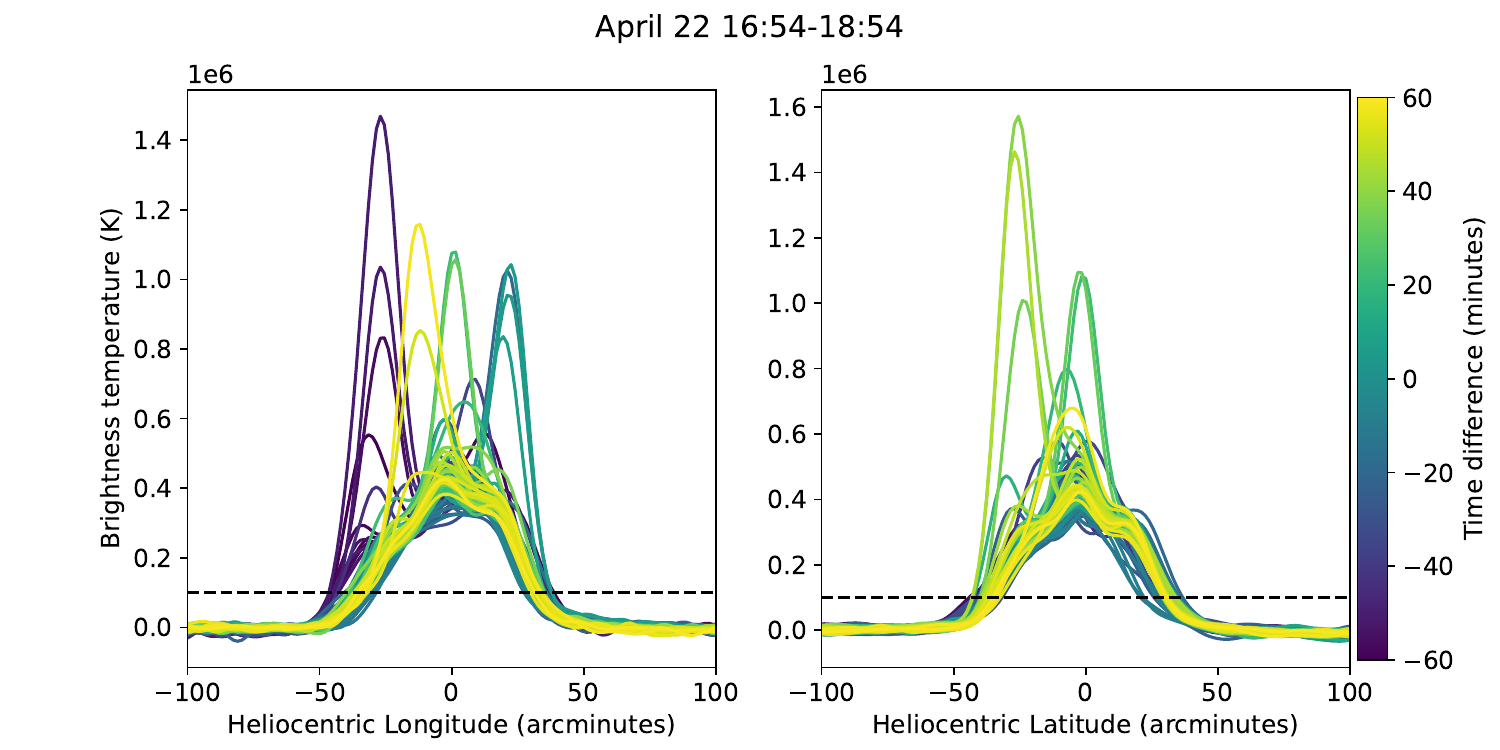}
    \includegraphics[scale=0.35]{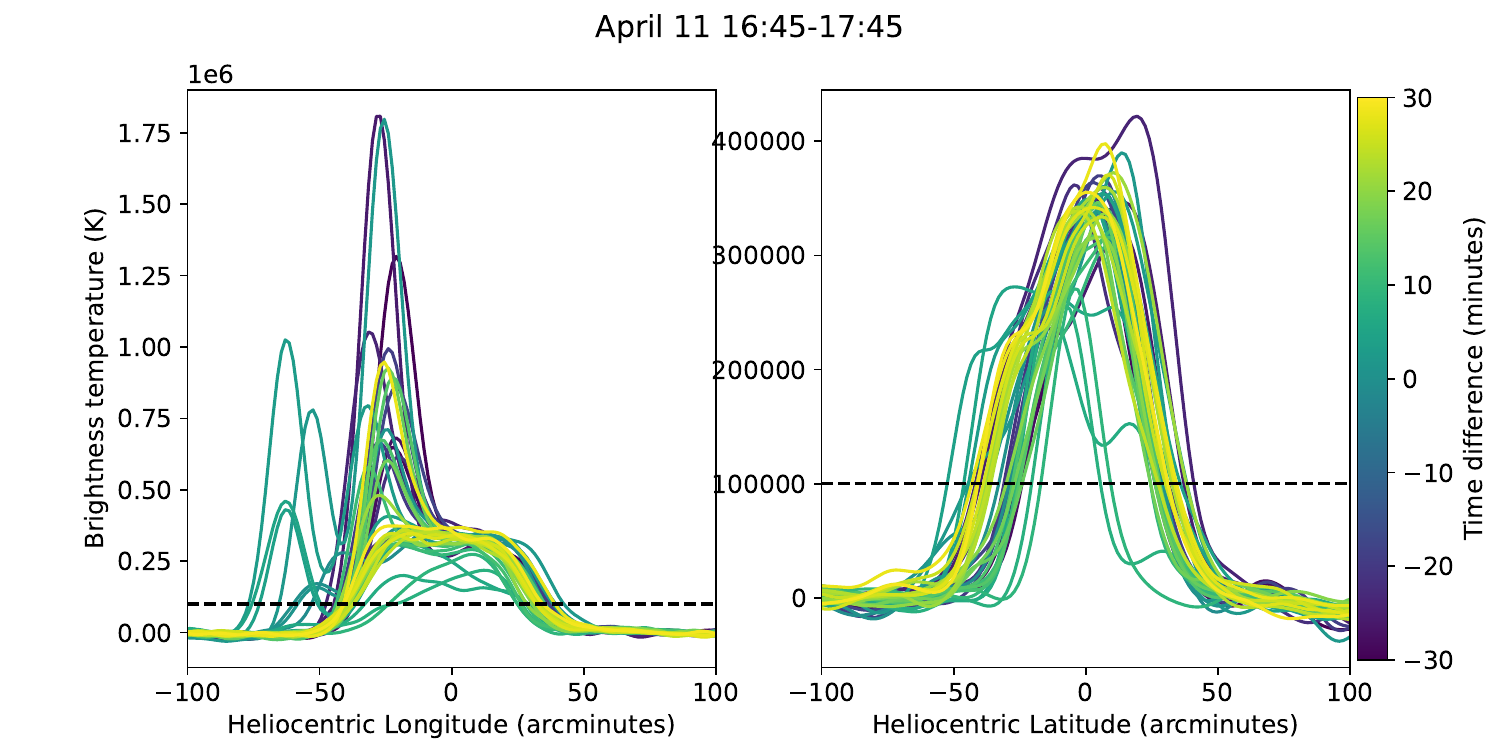}
    \caption{Panels from top left in clockwise manner: April 12, April 14, April 11 and April 22, 2024. The colorbar shows the time difference in minutes from the event time shown in Figures \ref{fig:20240412_event}, \ref{fig:20240414_event} and \ref{fig:20240422_event} for April 12, April 14 and April 22. April 22, shows a one hour window where we identified strong ionospheric activity. The black dashed line marks the 0.1 MK brightness temperature.{ The left and right subpanels of each day show the flux density variation along a 1D cut aligned along the heliocentric longitude and latitude respectively. The 1D cuts passes through the solar center on the sky plane. }}
    \label{fig:ionospheric_activity}
\end{figure*}

\subsection{Emission mechanism} \label{sec:emission_mechanism}

We consider three possible emission mechanisms for the observed extended weak radio transient structures: coherent plasma emission, thermal free-free emission, thermal gyroresonance emission, and nonthermal gyrosynchrotron emission. In the following subsections, we consider whether they can explain the observed emissions.

\subsubsection{Plasma emission} \label{sec:plasma_emission}

Plasma emission originates due to the interaction of nonthermal electrons and the ambient plasma. The emission happens at fundamental and harmonic of the local plasma frequency. To investigate if the emission can indeed originate due to plasma emission mechanism, we first determine the plasma density following \citet{hayes2001} using the total brightness maps obtained from LASCO C2. The LASCO C2 data were calibrated using the task \texttt{lasco\_prep} available in SolarSoft \citep{freeland1998,freeland2012}. In Figure \ref{fig:plasma_density} we show the obtained density distribution at times close to the radio images where we have detected the extended emissions. In the bottom right panel, we have shown a 1D distribution of the density at few chosen heliocentric distance, as a function of $\theta$, where $\theta$ is the angle from the solar north. {For determining the density from single line-of-sight images, as done here, one assumption typically made is that most of the mass, which is contributing to the observed white-light emission, is located on the plane of sky. \citet{vourlidas2000} investigated the effect of this assumption on mass estimates of CMEs, which is again directly related to the density estimates. They found that for CMEs with angular widths $\lesssim 60^\circ$ the mass is underestimated by $\approx 15\%$. For structures with the line-of-sight angle close to $20^\circ$, the mass is underestimated by about a factor of 5. }

\begin{figure*}
    \includegraphics[scale=0.5]{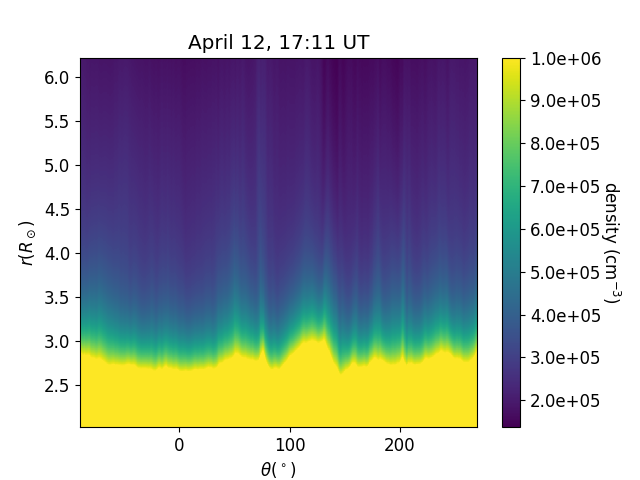}
    \includegraphics[scale=0.5]{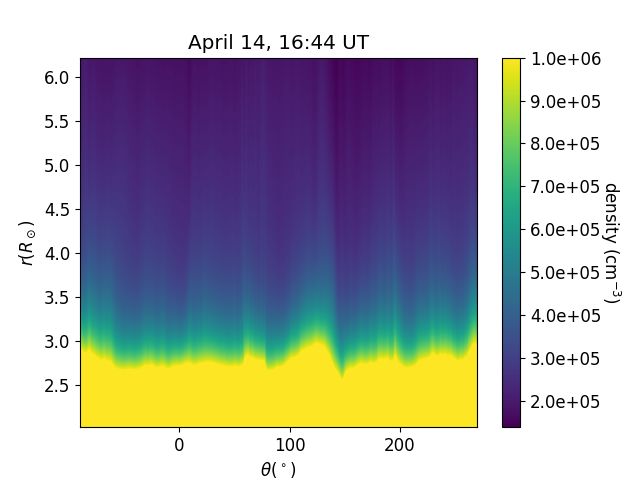}
    \includegraphics[scale=0.5]{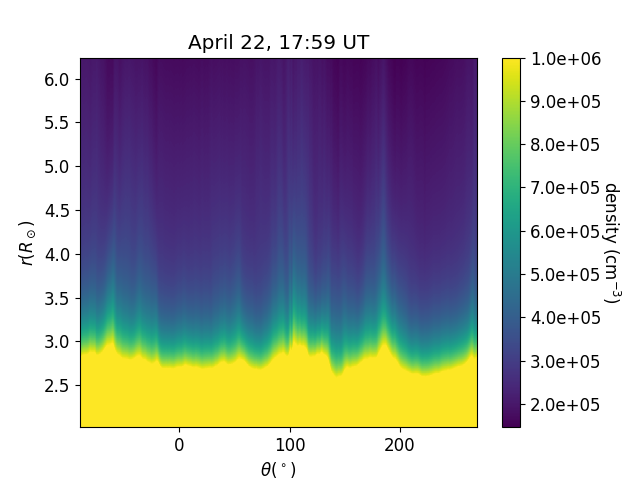}
    \hspace*{1.5cm}\includegraphics[scale=0.5]{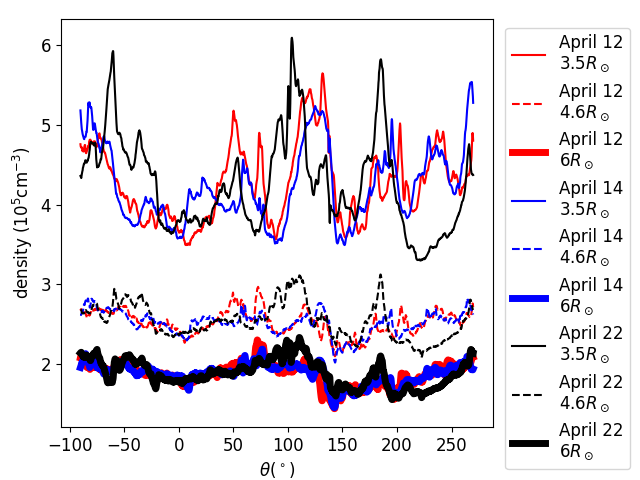}
    \caption{Top left, top right and bottom left panels show the density distribution as a function of heliocentric distance and position angle, measured from the solar north. The dates and times of the LASCO C2 images used to determine the densities are also shown in the respective titles. The bottom right panel shows a few 1D cuts at chosen heliocentric distances as a function of position angles at the three date-times shown in the other panels.}
    \label{fig:plasma_density}
\end{figure*}

On April 12 at 17:16 UT, the extended emission, as shown in Figure \ref{fig:20240412_event}, is seen from $\sim 3.3-6.5 R_\odot$, at a position angle of $\sim 100^\circ$. Using the density distribution shown in Figure \ref{fig:plasma_density}, we calculate that the plasma frequency at this position angle and heliocentric distance ranges from $6-4$ MHz. Thus, the plasma frequency is 7--10 times smaller than 43 MHz, where the emission was observed. Additionally, although the density decreased by more than a factor of 2 between 3--6.5 $R_\odot$, the emission is observed at the same frequency throughout this heliocentric distance. However, this can be explained by assuming the presence of an elongated high-density structure with a line-of-sight angle $\lesssim 25^\circ$ \citep{vourlidas2000} . {Density structures with a small line-of-sight angle will have very weak signature in the white light observations. The white-light observations when inverted regularly assume that most of the emission is arising from the sky plane, and thus will incorrectly assign very small densities to these density structures with small line of sight angles. }Additionally, such elongated heliocentric structures often do not show rapid density drops with heliocentric distance, and hence can explain the observed highly elongated radio sources. {Such high densities can also be observed if there are narrow high-density structures, with widths much smaller than the instrumental resolution of the white-light images used in this study. In the low corona, there exist multiple examples where high-density structures, observed in radio bands, show counterparts in the extreme ultraviolet and X-ray wavelengths \citep[e.g.][]{chen2013_typeIII}. Recent high-resolution observations from the Wide Field Imager for Solar Probe \citep[WISPR,][]{vourlidas2016} onboard the Parker Solar Probe have also discovered multiple small-scale, high-density features \citep{liewer2023}.
%The same arguments also hold true for other days as well.  
Thus, we believe that plasma emission due to nonthermal electrons is a plausible emission mechanism behind the observed radio emission. }

\subsubsection{Thermal Free-free emission} \label{sec:free_free}

To investigate if the observed emission can arise because of thermal free-free emission, we calculate the expected brightness temperature spectrum using the density estimated in Section \ref{sec:plasma_emission}. In Figures \ref{fig:20240412_event}, \ref{fig:20240414_event} and \ref{fig:20240422_event}, it is evident that the observed brightness temperature spectra are much steeper than the expected $\nu^{-2}$ behavior expected from optically thin free-free emission in absence of magnetic field. Such a steep spectrum is possible in presence of magnetic field, close to the gyro-frequency. We consider this possibility in Section \ref{sec:gyroresonance}. 
%The density chosen is greater than all the densities estimated at the location of the transient emissions. For April 12, the chosen density is larger than the observed density at $7.2 R_\odot$, even after considering an error of a factor of 5. Hence the predicted brightness temperature can be considered as an upper limit to the expected brightness temperature. 

\subsubsection{Thermal Gyroresonance emission} \label{sec:gyroresonance}

Thermal gyroresonance emission can arise due to thermal electrons gyrating in the plasma with an appreciable magnetic field. To understand the effect of the magnetic field in a quantitative manner, we have used the software implemented by \citet{fleishman2010}. In the left panel of Figure \ref{fig:predicted_free_free} we have shown the expected brightness temperature spectrum for varying magnetic field for a homogeneous plasma with density and temperature of $10^6$cm$^{-3}$ and 1 MK, respectively. We choose the depth along the line of sight to be 435 Mm. This is the same as the approximate width of the instrumental point-spread-function during this time, and essentially is the same as assuming a spherical source. 
We find that in the absence of magnetic field, the peak brightness temperature is less than $10^4$K, and hence is inconsistent with the observations. {If we assume an extremely high density of $5 \times 10^6$cm$^{-3}$, we find that the spectrum between 30--90 MHz is optically thin, with a frequency dependence of $\nu^{-2}$. This is also inconsistent with the observation.

In the presence of a magnetic field, we observe a sharp increase in the brightness temperature exceeding 0.1 MK at the gyrofrequency and a steep decline away from it. Thus, thermal gyroresonance emission can explain the observed brightness temperatures and the steep spectra. Under the assumption that the emission is produced by thermal plasma, the transient radio emission can only be explained if new magnetized plasma is present at the location of the radio source. While the presence of a magnetic field cannot be tested using other means, we test this hypothesis by studying the time-variable plasma density at the locations of the radio source. 

In the left panel of Figure \ref{fig:density_diff}, we show the density difference at 17:59 UT and 17:47 UT on April 22, 2022. These times are the 2 nearest LASCO C2 images available to the time shown in Figure \ref{fig:20240422_event}. In the right panel, we show the density difference at a heliocentric distance of 3.75$R_\odot$. We find that at 3.75 R$_\odot$, the maximum density difference is about $2\times 10^4$cm$^{-3}$. On this date, the maximum density difference is seen, out of all the times considered here. This density difference is $\lesssim 0.06$ times the density at these heights (bottom right panel of Figure \ref{fig:plasma_density}). This small density difference is also consistent with the fact that we do not detect significant white-light emission in any of the locations where we have detected these extended radio emissions. We now simulate the condition that a small amount of magnetized plasma has been ejected from the sun and is present during the time of detection at the location of the radio source. This inhomogeneous configuration can also be simulated by the software implemented in \citet{fleishman2010}. In the right panel of Figure \ref{fig:predicted_free_free} we show the result of this simulation, where we have varied the fraction of magnetized plasma. The magnetic field is assumed to be 10 G. Hence, the gyrofrequency is 28 MHz, which is outside our observation range. Since the frequency range studied here falls in the optically thin regime, the location of the magnetized plasma along the line-of-sight (LOS) is not important.  We have arbitrarily set the location of the magnetized plasma to be located at the centre of the LOS. The density of the plasma is set to $4\times10^5$cm$^{-3}$, { which is 20 times greater than the density difference observed in Figure \ref{fig:density_diff} at 3.75 R$_\odot$.}
%which is the typical density at $\sim 3.5 R_\odot$. 
All other parameters are same as those in the left panel of Figure \ref{fig:predicted_free_free}. It is evident that the fraction of the magnetized plasma should be close to 100\% to explain the observed brightness temperature. Since the magnetised plasma is primarily contributing to the observed brightness temperature peak of $\approx 10^5$K, very high density plasma, comparable to the background plasma density at $\sim 3 R_\odot$, should be repeatedly ejected from the Sun, to make the radio observations possible. Significant variation in the magnetic field within the magnetized plasma blob is also allowed as that is likely to create a much flatter spectrum, which is again inconsistent with observations. Additionally, it should be quite collimated along the LOS, as we do not see evidence of such strong density enhancement in the white-light and also move with high velocity to explain its transient nature in the radio data. {We can estimate the minimum velocity required using the images shown in Figures \ref{fig:20240412_time_variability}, \ref{fig:20240414_event} and \ref{fig:20240422_time_variability}. In Figure \ref{fig:20240412_movie} we find that the transient emission is not detected at 16:51 UT, but is detected at a sky coordinate of approximately ($-6000\arcsec,-4000\arcsec$) at 16:57 UT. This implies that if a magnetized plasma blob is responsible for the observed radio emission, it should cover a distance of $\sim 4.6R_\odot$ in the sky plane in 6 minutes. This translates to a a plane-of-sky velocity of 8900 km/s . Similar arguments, when applied to Figures \ref{fig:20240414_event} and \ref{fig:20240422_time_variability}, also lead to velocities exceeding 4000 km/s. The inferred velocities is much larger than the fastest CME detected till date. Based on this, we believe that while theoretically, { gyroresonance} emission due to a high-velocity dense blob of homogeneous magnetized plasma can explain the observations, it is very unlikely. }

%The highest velocity requirement is on April 14, where the emission is not detected after 3 minutes, even though the noise in the images looks similar. Using this information we estimate that the velocity of the plasma should be $\sim 4000$km/s (assuming a plane of sky motion of $\sim 1000\arcsec$ in 3 minutes), which is much faster than the fastest CME detected yet. 
%Since free-free emission increases with increase in density, since the maximum density difference was unable to explain the observed transient emission, the same will hold true for other times shown here as well. We note that that it is possible that LASCO C2 was unable to observe the density enhancement because of limited temporal resolution. However, based on typical velocities of plasma blobs, we do not except the blob to move outside the LASCO C2 field of view within the cadence of white light images and should have been seen in the density differences if present. 

\begin{figure*}
    \centering
    \includegraphics[scale=0.5]{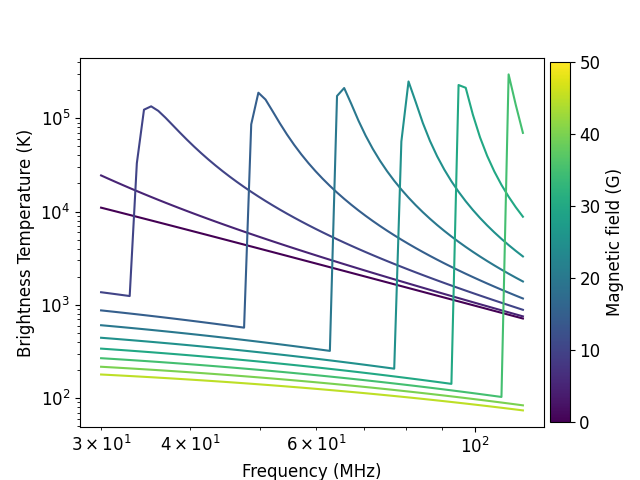}
    \includegraphics[scale=0.5]{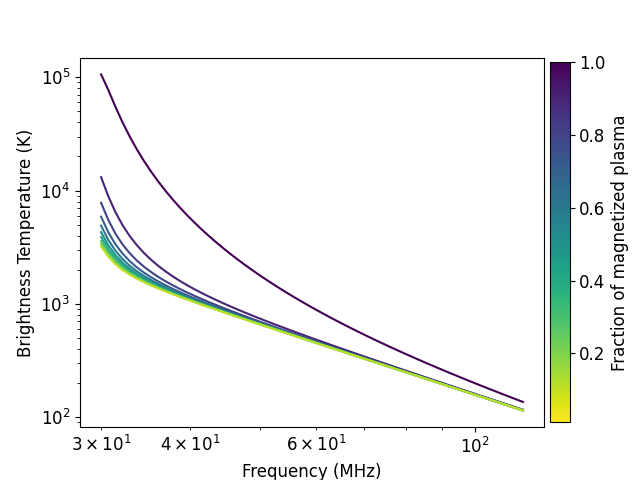}
    \caption{Left panel: Shows the expected free-free emission for a 1 MK plasma of density $10^6$cm$^{-3}$ in presence of different magnetic field strengths. Right panel: Predicted brightness temperature for varying fraction of magnetized plasma due to combination of gyroresonance and thermal free-free emission. The magnetic field is assumed to be 10G.}
    \label{fig:predicted_free_free}
\end{figure*}

\begin{figure*}
    \centering
    \includegraphics[scale=0.55]{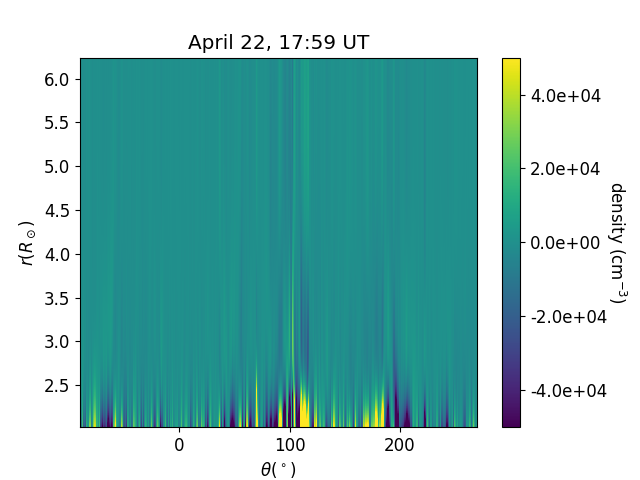}
    \includegraphics[scale=0.5]{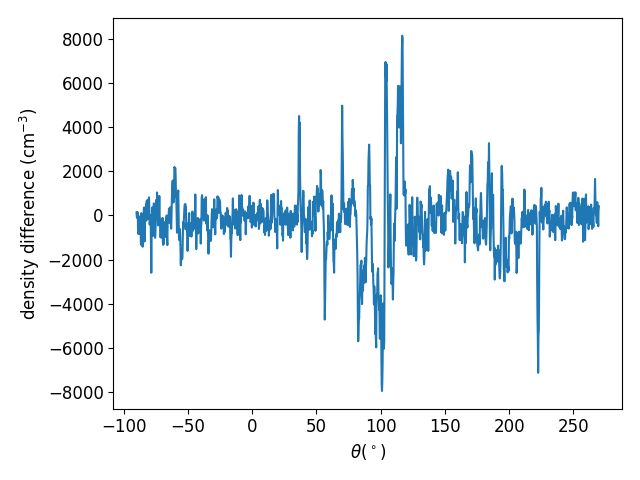}
    \caption{Left panel: Density difference between April 22 17:59 and 17:47 UT as a function of heliocentric distance and position angle, measured from solar north. The colorscale has been saturated between $-4\times 10^4$ and $4\times 10^4$cm$^{-3}$. Right panel: Shows the density difference at 3.75 $R_\odot$ as a function of position angle. }
    \label{fig:density_diff}
\end{figure*}

% \begin{figure}
%     \centering
%     \includegraphics[scale=0.5]{predicted_brightness_temp_free_free_inhomogeneous_10G.png}
%     \caption{Predicted brightness temperature for varying fraction of magnetized plasma due to combination of gyroresonance and thermal free-free emission. The magnetic field is assumed to be 10G.}
%     \label{fig:inhomogeneous_free_free}
% \end{figure}

\subsubsection{Gyrosynchrotron emission} \label{sec:gyro_emission}

Gyrosynchrotron emission is an attractive option to explain this phenomenon. Nonthermal electrons can easily spread throughout the radio-emitting region and produce gyrosynchrotron emission. Gyrosynchrotron emission with similar brightness temperatures from similar heliocentric distances has been reported in the literature \citep{mondal2020, kansabanik2023}. However, they were associated with CMEs, unlike the case here. However, the possibility of gyrosynchrotron emission powering the observed radio structures is proven by past works. {As an illustration, in Figure \ref{fig:april12_20240412_GS}, we have shown a model gyrosynchrotron spectrum, which can explain the observed data from April 12 (Figure \ref{fig:20240412_event}). We have used a $\kappa$ distribution to model the electron distribution. The relevant parameters are indicated in the figure. $\eta$ is the filling factor and is the ratio between the emission area and the area covered by the beam when projected onto the Sun. $B$, $\theta$, $n_e$, $L$ are the magnetic field, angle of the magnetic field with the LOS, electron density, and depth along LOS, respectively. $\kappa$ is the index of the $\kappa$ distribution. It should be noted that the model spectrum shown here is only for demonstration that the observed data can be explained using a gyrosynchrotron emission mechanism.} %While the model can describe the spectrum quite nicely, we are not confident about the parameters, and hence we have chosen to only say that the GS model can be tuned to explain the spectra, and have not provided any quantitative model. "Detailed spectral fitting" in our opinion, implies that the parameters determined from the fitting are reliable and represent the "reality" to a significant extent. In this case, we are not confident that this is indeed the case.

 { We ruled out gyroresonance emission based on the high velocity requirements on the magnetized plasma blob. However the velocities required ($\sim 8000$km/s) is only 0.02c. Nonthermal electrons with speeds much greater than this value are routinely observed in the solar corona and heliosphere. The short lifetimes ($\sim$a few minutes) can also be explained using nonthermal electrons. If the responsible non-thermal electrons are free-flowing, then their often observed mildly relativistic nature will allow them to move out of the observed emitting region in a very small time interval. This implies, in this case, there needs to be a continuous supply of nonthermal electrons throughout the duration of the observed radio emission. The transient nature of the observed radio emission can be due to the transient nature of the particle acceleration processes happening in the lower corona. On the other hand, if the nonthermal electrons can get trapped in the coronal structures, then the observed emission lifetime would be dependent on the trapping timescale. Thus, in this scenario, while there needs to be multiple episodes of nonthermal particle injection, the particle acceleration duration of each episode can be much smaller than the few minutes emission lifetime observed here. Although our data do not allow us to perform a detailed spectral fitting, we have shown that the distribution of nonthermal electrons and magnetic field, the two key quantities for determining the observed brightness temperature of gyrosynchrotron emission, can be tuned to qualitatively match the observed radio emissions.}

 \begin{figure}
     \centering
     \includegraphics[scale=0.5]{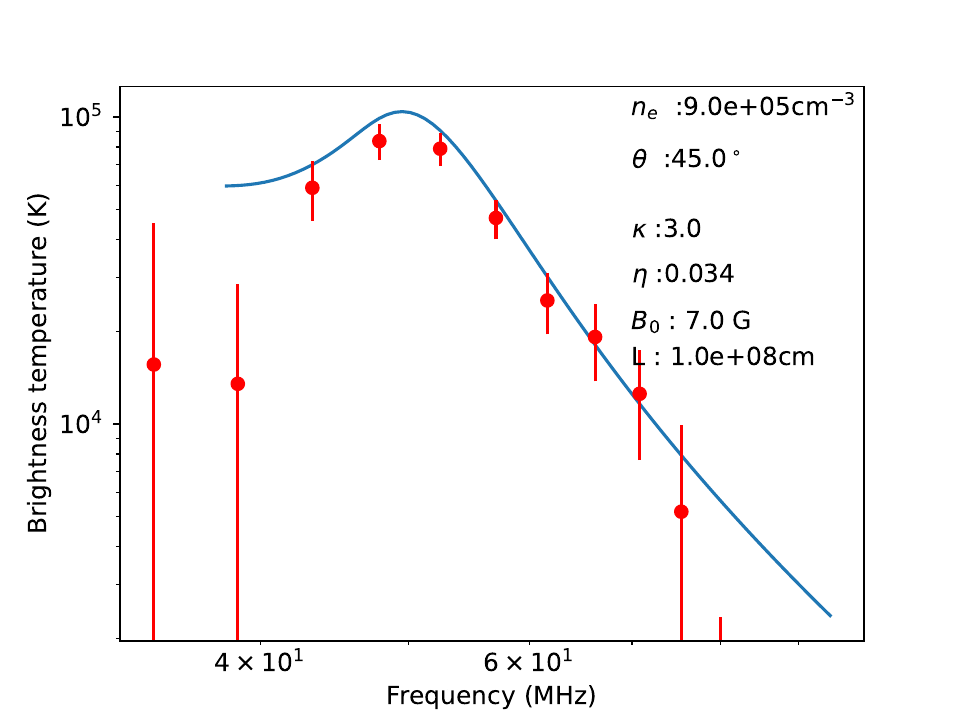}
     \caption{Shows a model gyrosynchrotron spectrum, which can explain the observed data from April 12 (Figure \ref{fig:20240412_event}).  The relevant parameters used to compute the model spectrum are indicated in the figure. }
     \label{fig:april12_20240412_GS}
 \end{figure}

\section{Conclusion} \label{sec:conclusion}

In this work, we have provided robust evidence of the presence of faint, transient extended emissions in the middle corona. These emissions are detected at large heliocentric distances exceeding $3 R_\odot$. In one instance, the emission extended to $\sim 7 R_\odot$, which is approximately the edge of the field of view of the image stored in the OVRO-LWA data archive. The lifetime of these emissions is found to be around a few minutes. In contrast to all past expectations, we do not find any significant white-light variability at the locations of these radio transients. { While this can be due to the limited cadence { and sensitivity} of available white-light observations, the sensitive radio observations presented here are bringing into light probably a new class of coronal transients, which has been hitherto unknown.} We have examined the probable emission mechanisms of these transient emissions in great detail. Based on our analysis, we find that these emissions are most likely to be powered by nonthermal electrons, such as incoherent gyrosynchrotron emission and coherent plasma emission. Hence, this study indicates the presence of nonthermal electrons in the middle corona, even in the absence of CMEs, and energetic flares. Type III radio bursts are often used to probe nonthermal electron beams.  However, the brightness temperatures of some of the weakest of type IIIs reported earlier have a brightness temperature of a few MK, more than an order of magnitude brighter than the brightness temperatures observed here. This suggests that the nonthermal electron population responsible for the radio emissions reported in this work, is probably much weaker than those responsible for the larger radio bursts, like type IIIs and type IIs associated with flares and CMEs.  It is also interesting to note that a few earlier low-frequency radio observations  \citep{mondal2020_winqse,sharma2022,mondal2023_ayan_data, mondal2023_spectrum} have also indicated the presence of rather weak nonthermal electrons even in the quiet solar corona. However, the origin of these nonthermal electrons is still not clear. Regular monitoring of the middle corona with high dynamic range low-frequency radio images with instruments like OVRO-LWA is crucial for answering these questions and for a detailed understanding of the dynamics of the middle corona and the Sun in general.

%In the past all detections of gyrosynchrotron emission in the middle corona corresponded to a CME. Hence this is the first suggestion of gyrosynchrotron emission in the middle corona in absence of a CME. If the emission is powered by gyrosynchrotron emission, we also conclude that these emissions are most probably powered by free-streaming nonthermal electrons or extremely low energy trapped nonthermal electrons. 

\begin{acknowledgments}
S.M., B.C., and D.G. acknowledge support by the NASA Living With a Star (LWS) Science grant 80NSSC24K1116. S.Y. was supported by the NASA Early Career Investigator Program (ECIP) grant to NJIT (80NSSC21K0623).
P. Z. acknowledges support for this research by the NASA Living with a Star Jack Eddy Postdoctoral Fellowship Program, administered by UCAR’s Cooperative Programs for the Advancement of Earth System Science (CPAESS) under award 80NSSC22M0097. The OVRO-LWA expansion project was supported by NSF under grant AST-1828784. OVRO-LWA operations for solar and space weather sciences are supported by NSF under grant AGS-2436999. A portion of this research was carried out at the Jet Propulsion Laboratory, California Institute of Technology, under a contract with the National Aeronautics and Space Administration (80NM0018D0004).  A portion of this work was supported by the National Science Foundation Graduate Research Fellowship under Grant No. 2139433. The authors also thank the referee for the comments, which have helped to improve the manuscript.
\end{acknowledgments}

\appendix

{In movies named 20240412\_movie.mp4 and 20240422\_movie.mp4 and \ref{fig:20240412_movie} and \ref{fig:20240422_movie}, we show a composite movie of radio and white light images (available in the online version) spanning the observation times presented here for April 12 and April 22. In each frame of the movies, we have shown multi-frequency radio images over a white light difference image. The difference is taken with respect to 15:16 and 14:18 for April 12 and April 14, respectively. The title of each panel shows the time of the radio image. The time written inside each panel shows the time of the background white-light image. Due to dynamic range considerations, some frames have fewer frequencies compared to other frames.}

\begin{figure}
    %\centering
    \begin{interactive}{animation}{20240412_movie.mp4}
    \includegraphics[trim={3cm 0 0 0},clip, scale=0.5]{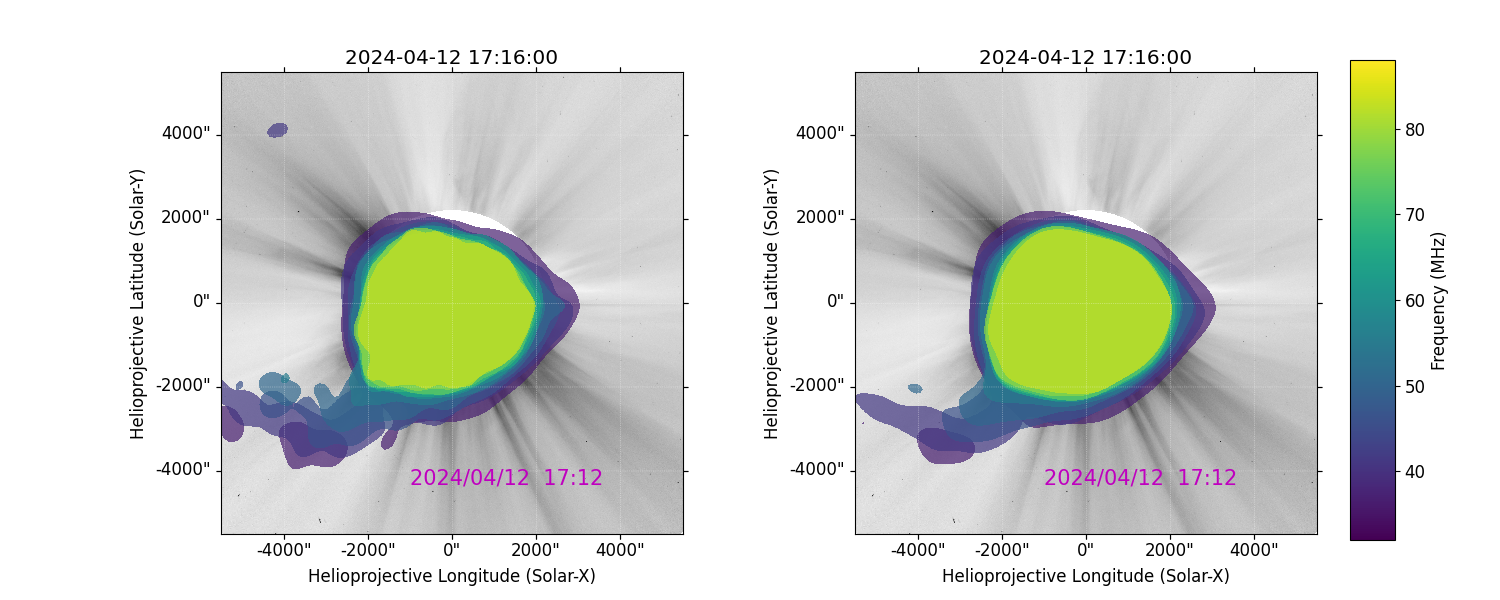}
    \end{interactive}
    \caption{Multi-frequency radio contours between 0.05 and 2 MK on the nearest available LASCO C2 difference image. The right panel shows radio images smoothed using a Gaussian function of standard deviation of $4.5\arcmin$. The left panel shows images before smoothing. The difference is taken with respect to the white light image at 15:16 UT. The time of the radio image is shown in the title of the figure. The observation time of the LASCO C2 image is also indicated with an annotation.} %10\% and 3\% of the peak in the respective images. }
    \label{fig:20240412_movie}
\end{figure}

\begin{figure}
    %\centering
    \begin{interactive}{animation}{20240422_movie.mp4}
    \includegraphics[trim={3cm 0 0 0},clip, scale=0.5]{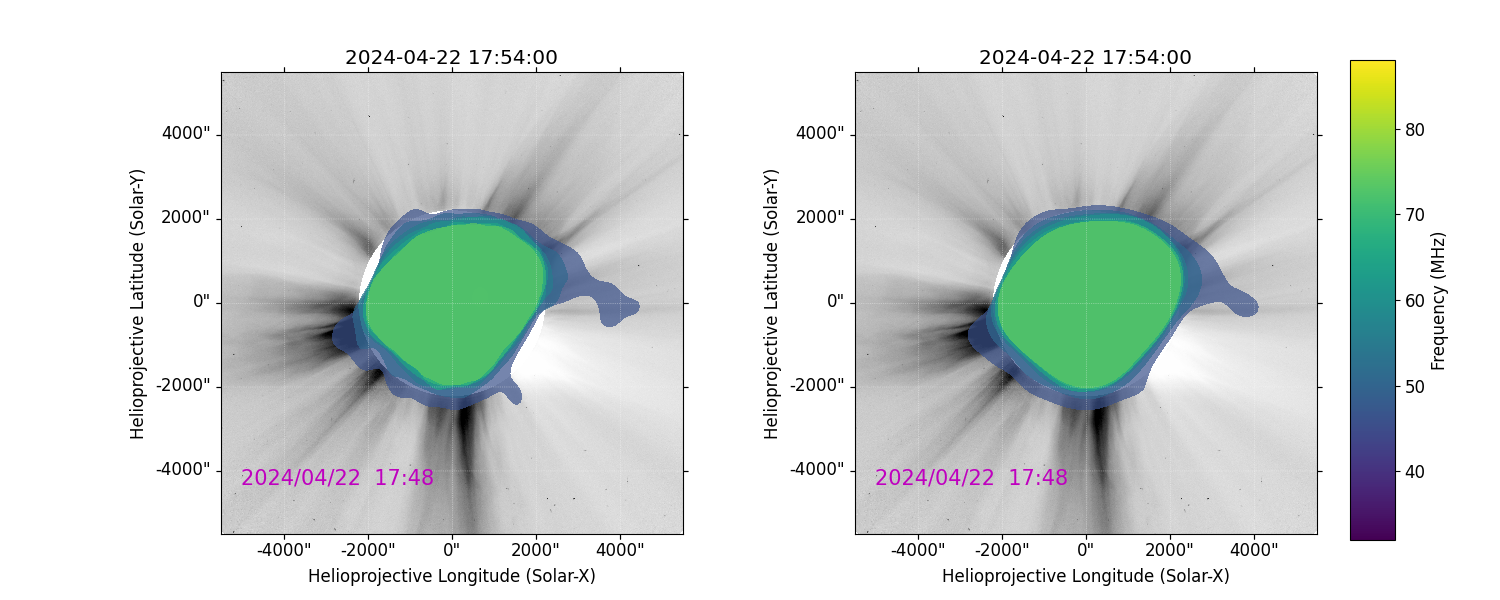}
    \end{interactive}
    \caption{Multi-frequency radio contours between 0.06 and 2 MK on the nearest available LASCO C2 difference image. The left panel shows radio images smoothed using a Gaussian function of standard deviation of $4.5\arcmin$. The right panel shows images before smoothing.  The difference is taken with respect to the white light image at 14:18 UT. The time of the radio image is shown in the title of the figure. The observation time of the LASCO C2 image is also indicated with an annotation. Due to dynamic range limitations, we have only plotted contours if the lowest contour level is detected at a signal-to-noise (SNR) ratio of 3. We have also limited ourselves to frequencies between 39 and 75 MHz, as some higher and lower frequencies were severely dynamic range limited at some times, and the chosen SNR threshold was not sufficient to remove them.} %10\% and 3\% of the peak in the respective images. }
    \label{fig:20240422_movie}
\end{figure}

\bibliography{bibliography}

\end{document}